\documentclass{article}
\newcommand{\npatients}{60,529} 
\newcommand{\nsamples}{67,608} 
\newcommand{\nslides}{75,383} 

\newcommand{\nprimary}{46,650}
\newcommand{\nmetastasis}{20,958}

\newcommand{\npatientstrain}{41,468}
\newcommand{\nsamplestrain}{46,201}
\newcommand{\nslidestrain}{51,747}

\newcommand{\npatientstrainmsk}{34,349}
\newcommand{\nsamplestrainmsk}{39,082}
\newcommand{\nslidestrainmsk}{43,022}

\newcommand{\npatientstraintcga}{7,119}
\newcommand{\nsamplestraintcga}{7,119}
\newcommand{\nslidestraintcga}{8,725}

\newcommand{\nsamplestune}{9,677}
\newcommand{\nslidestune}{10,710}

\newcommand{\nsamplestest}{9,862}
\newcommand{\nslidestest}{10,645}

\newcommand{\nsamplestestcommoncancers}{8,178}
\newcommand{\nsamplestestcommonprimary}{5,260}
\newcommand{\nsamplestestcommonmetastasis}{2,918}

\newcommand{\nslidestcgaoriginal}{11,406}

\newcommand{\nsamplestcgaeval}{1,868}

\newcommand{\nslidestcgaeval}{2,281}

\newcommand{\nvectorlabels}{1,637}
\newcommand{\nimpactgenes}{489} 

\usepackage{authblk}
\usepackage[printonlyused, nohyperlinks, nolist]{acronym}
\usepackage{afterpage}
\usepackage{amsmath}
\usepackage[english]{babel}
\usepackage{booktabs}
\usepackage{caption}
\usepackage{cite}
\usepackage{comment}
\usepackage{graphicx} 
\usepackage{hyperref} 
\usepackage{listings}
\usepackage{longtable}
\usepackage{makecell}
\usepackage{mleftright}
\usepackage{multirow}
\usepackage{pdflscape}
\usepackage{placeins}
\usepackage{subcaption}
\usepackage{tikz}
\usepackage[nottoc]{tocbibind}
\usepackage{tabularx}
\usepackage{url} 
\usepackage{xcolor}

\DeclareCaptionLabelFormat{subtablelabelformat}{\tablename\ #2}
\makeatletter
\renewcommand\p@subtable{}
\makeatletter

\definecolor{codegreen}{rgb}{0,0.6,0}
\definecolor{codered}{rgb}{1,0,0}
\definecolor{codegray}{rgb}{0.5,0.5,0.5}
\definecolor{codepurple}{rgb}{0.58,0,0.82}
\definecolor{backcolour}{rgb}{0.95,0.95,0.92}

\lstdefinestyle{mystyle}{
  backgroundcolor=\color{backcolour},
  commentstyle=\color{codegreen},
  keywordstyle=\color{magenta},
  numberstyle=\tiny\color{codegray},
  stringstyle=\color{codepurple},
  basicstyle=\ttfamily\footnotesize,
  breakatwhitespace=false,         
  breaklines=true,                 
  captionpos=b,                    
  keepspaces=true,                 
  numbers=left,                    
  numbersep=5pt,                  
  showspaces=false,                
  showstringspaces=false,
  showtabs=false,                  
  tabsize=2,
  escapeinside={(*@}{@*)}
}

\DeclareCaptionSubType*[arabic]{table}

\title{Screen Them All: High-Throughput Pan-Cancer Genetic and Phenotypic Biomarker Screening from H\&E Whole Slide Images}
\author[1,4]{Yi Kan Wang}
\author[1,4]{Ludmila Tydlitatova}
\author[1,4]{Jeremy D. Kunz}
\author[1]{Gerard Oakley}
\author[1]{Bonnie Kar Bo Chow}
\author[1]{Ran A. Godrich}
\author[1]{Matthew C. H. Lee}
\author[1]{Hamed Aghdam}
\author[1]{Alican Bozkurt}
\author[1]{Michal Zelechowski}
\author[2]{Chad Vanderbilt}
\author[3]{Christopher Kanan}
\author[1]{Juan A. Retamero}
\author[1]{Peter Hamilton}
\author[1]{Razik Yousfi}
\author[1]{Thomas J. Fuchs}
\author[1]{David S. Klimstra}
\author[1,5]{Siqi Liu}
\affil[1]{Paige, NYC, NY United States}
\affil[2]{Memorial Sloan Kettering Cancer Center, NYC, NY United States}
\affil[3]{University of Rochester, Rochester, NY United States}
\affil[4]{Equal contribution}
\affil[5]{Corresponding author (siqi.liu@paige.ai)}
\date{}

\begin{document}

\setcounter{figure}{0}
\renewcommand{\thefigure}{\arabic{figure}}

\raggedbottom

\maketitle
\begin{abstract}

Molecular assays are standard of care for detecting genomic alterations in cancer prognosis and therapy selection but are costly, tissue-destructive and time-consuming.
Artificial intelligence (AI) applied to routine \ac{HE}-stained \acp{WSI} offers a fast and economical alternative for screening molecular biomarkers.
We introduce OmniScreen, a high-throughput AI-based system leveraging Virchow2 embeddings extracted from \npatients{} cancer patients with paired \nimpactgenes{}-gene MSK-IMPACT targeted biomarker panel and WSIs. 
Unlike conventional approaches that train separate models for each biomarker, OmniScreen employs a unified model to predict a broad range of clinically relevant biomarkers across cancers, including low-prevalence targets impractical to model individually.
OmniScreen reliably identifies therapeutic targets and shared phenotypic features across common and rare tumors. We investigate the biomarker prediction probabilities and accuracies of OmniScreen in relation to tumor area, cohort size, histologic subtype alignment, and pathway-level morphological patterns. These findings underscore the potential of OmniScreen for routine clinical screening.
\end{abstract}

\section{Introduction}

Single and multi-gene assays have revolutionized personalized cancer treatments by identifying targetable genomic alterations~\cite{passaro2024cancer,zhou2024tumor,beyer2022diagnostic}. For example, \ac{AR} variants predict endocrine versus chemotherapy resistance in metastatic castrate resistant prostate cancer (mCRPC)~\cite{graf2022predictive}, while BRCA1/2 germline mutations guide \ac{PARP} inhibitor use in the treatment of high-risk early stage HER2-negative breast cancer~\cite{tutt2021adjuvant}. Commercial assays such as Oncotype DX, MammaPrint and Decipher provide valuable prognostic insights but are costly, time-consuming, and require substantial tissue samples, limiting their utility in cases with scarce tumor material~\cite{eggener2020molecular}. 
Minimally invasive approaches, such as liquid biopsies, have emerged to address these issues
but face challenges including low sample yield, variability in cell collection and stabilization procedures, and reduced sensitivity and specificity due to the presence of non-tumor mutated clones in the blood of patients without cancer~\cite{habli2020circulating,cirmena2021assessment}

Digital biomarkers derived from routine \ac{HE} \acp{WSI} may sidestep many of these limitations. By leveraging widely available WSIs and digital image analysis, such biomarkers could enable rapid, cost-effective and tissue-sparing screening, especially if clinically important genomic biomarkers can be identified. Such applications could complement existing workflows by prioritizing cases for definitive genomic testing while excluding those unlikely to yield actionable results, thereby improving efficiency in both cost and turnaround time.
Beyond clinical applications, digital biomarkers could accelerate drug development and clinical trials by facilitating novel target discovery, optimizing patient stratification, and improving the cost efficiency and success rates of clinical studies~\cite{ramon2024development,vamathevan2019applications}. Moreover, robust digital biomarkers could eventually serve as companion diagnostics in clinical practice, integrating seamlessly into precision oncology workflows.

Recent studies have demonstrated the ability of machine learning methods to identify morphological features in routine \ac{HE} histopathology images that are associated with genomic abnormalities across various cancer types, enabling the prediction of clinically actionable genomic alterations directly from WSI analysis~\cite{fu2020pan,kather2020pan,qu2021genetic,volinsky2024prediction,campanella2022h,arslan2024systematic}. However, most of these efforts focused on single biomarkers for specific tissues or cancer types, limiting their generalizability due to restricted training data and overlooking shared morphological patterns across cancers. 
Additionally, developing separate models for each biomarker is both resource-intensive and inefficient.
Despite advances in digital pathology, the simultaneous identification of multiple biomarkers across diverse tumor types remains limited, largely due to insufficient access to genomically characterized WSIs. Existing datasets, such as \ac{TCGA}, have been widely used for multi-biomarker studies but were primarily designed for genomic research, with digital image submission as a secondary priority~\cite{weinstein2013cancer}. Variability in image quality and sequencing methods may impact model performance and introduce biases in predictive accuracy~\cite{howard2021impact,couture2022deep}.

\begin{figure}[!ht]
    \centering
    \includegraphics[width=\textwidth]{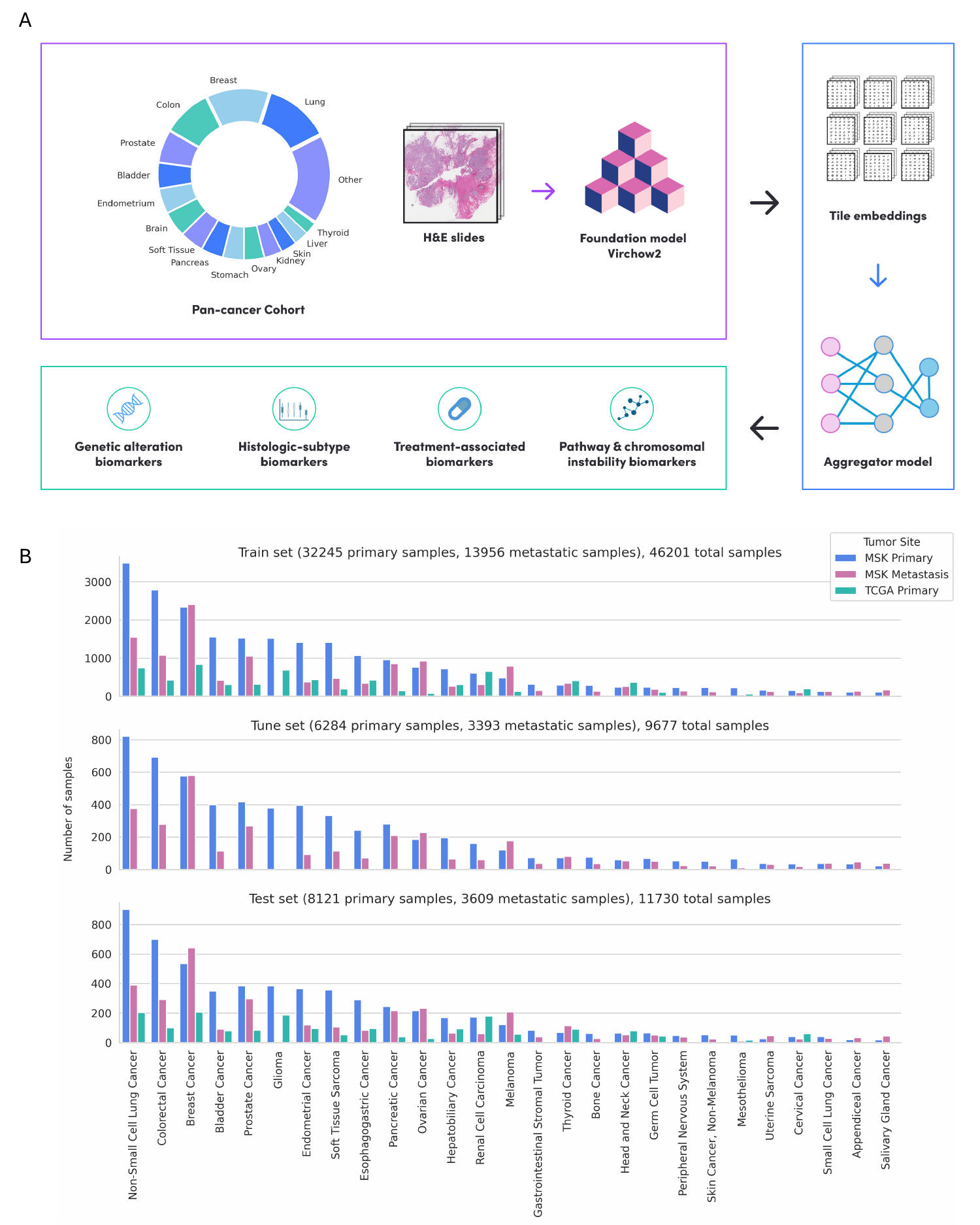}
    \caption{Pan-cancer digital biomarker identification from \ac{HE} \acfp{WSI}. 
    \textbf{A}: The OmniScreen workflow: \ac{HE} WSIs from diverse tumor histologies are processed using Virchow2 to generate tile embeddings. These representations, along with genomic ground truth, serve as input to train an aggregator network for biomarker prediction. The use cases of biomarker predictions include genetic alteration prediction, genotype-phenotype correlation, treatment target prediction, and pathway and chromosomal instability phenotype detection. 
    \textbf{B}: Sample counts across diverse cancer types in the dataset, split by train, tune and test. Remaining rare cancers account for 4\% of all samples (category `Other', not visualized). 3\% of all samples do not have an associated cancer type (category `Unknown', not visualized). Sample distributions are stratified by data cohort and tumor site.} 
    \label{fig:1}
\end{figure}

Here, we present an approach to high-throughput screening for clinically relevant genomic abnormalities applicable to all cancer types using routine \ac{HE}-stained \aclp{WSI} (Figure~\ref{fig:1}). By leveraging image representations from Virchow2, a foundation model pre-trained on 3 million slides~\cite{vorontsov2024foundation,zimmermann2024virchow2scalingselfsupervised}, we develop OmniScreen, a model capable of simultaneously predicting \nvectorlabels{} biomarkers across 71 human cancers. OmniScreen was trained and tested on a cohort of \nslides{} \acp{WSI} obtained from \npatients{} patients, sourced from \ac{MSKCC} and \ac{TCGA}. MSKCC is a large cohort of WSIs with associated comprehensive molecular characterization, with single institution control of tissue processing, image generation, and molecular characterization using the \ac{FDA}-approved \ac{MSK-IMPACT} \ac{NGS} panel. This eliminates many of the limitations of other public databases. Inclusion of TCGA ensures generalizability to other tissue processing methods, staining procedures, and molecular characterization methods.

We evaluated our model across diverse tumor types, identifying  clinically relevant biomarkers associated with driver genomic alteration, histologic phenotypes, actionable targets for FDA-approved therapies, molecular pathways and genome instability. We further examined the impact of training sample size and tumor area within WSIs on biomarker prediction. Finally, we compared OmniScreen to the conventional methods that involve training separate models for individual biomarkers. By enabling scalable and accurate predictions from \ac{HE}-stained WSIs, our model demonstrates the potential to complement molecular assays, refine prognostic assessments, guide therapeutic strategies, and address challenges in integrating image-based biomarkers into clinical oncology.

\section{Results}

\subsection{OmniScreen enables high-throughput pan-cancer biomarker screening}

Our model, OmniScreen, simultaneously predicts the likelihood of \nvectorlabels{} biomarkers, representing genomic features derived from the MSK-IMPACT panel of \nimpactgenes{} cancer-related genes. These biomarkers include pathogenic alterations, measures of genomic instability, and alterations in groups of genes involved in shared signaling pathways (Figures~\ref{fig:S1},~\ref{fig:S2}, Methods~\ref{method:model}). 


We evaluated biomarker performance across tumor types in both primary and metastatic sites (Figures~\ref{fig:S3}A-B). To address different use cases, we defined filtering criteria to categorize biomarkers into tiers, as outlined in Table~\ref{table:filter-criteria} (see Methods~\ref{section:supplementary-methods-label-filters}).
The baseline criteria identified 739 genetic alteration biomarkers (AUC~$>0.5$), spanning 346 genes in 43 cancer types, exceeding the threshold expected by chance. Notably, the accuracy of biomarker predictions varied across cancer types (Figure~\ref{fig:2}A).

Using high-performing biomarker criteria, we further refined this set to include genomic alterations with strong screening potential for the 15 most common cancer types treated at MSK. From \nsamplestestcommoncancers{} samples (\nsamplestestcommonprimary{} primary and \nsamplestestcommonmetastasis{} metastatic samples), we identified 427 biomarkers with a mean AUC of 0.84 (mean sensitivity=0.92 and mean specificity=0.55). This refined set highlights the promise for screening 235 genes in at least one cancer type (Table~\ref{table:msk-evaluation}). We observed higher AUCs when evaluating WSIs obtained from primary tumors (mean AUC 0.86; range 0.76-0.96) compared to those from metastatic lesions (mean AUC 0.82; range 0.77-0.95) (Figure~\ref{fig:S3}C-D, Table~\ref{table:msk-evaluation}). Notably, strong predictive performance was evident for genomic alterations in well-established oncogenes and tumor suppressors, such as TP53, CDKN2B, TERT, CDKN2A, RB1, MYC, PTEN (Figure~\ref{fig:S3}E). Among tumor types, \acp{CRC} demonstrated the broadest range of biomarkers with strong screening potential, achieving AUC~$>0.75$ for 80 genes in primary \acp{CRC} and 29 genes in metastatic \acp{CRC}. Endometrial cancers followed, exhibiting similar robust performance, with AUC~$>0.75$ for 50 different genes in primary tumors and 28 genes in metastatic samples (Figure~\ref{fig:S3}F).

Top-performing biomarkers, defined as those with prevalence~$>5\%$ and AUC~$>0.85$ (Table~\ref{table:filter-criteria}), comprise 93 genomic alterations among 55 genes. These biomarkers achieved a mean AUC of 0.90 (range: 0.85-0.99), with a mean sensitivity of 0.93, and a mean specificity of 0.66 (Figure~\ref{fig:2}B).
Of these, 29 genomic alterations which were sufficiently represented in the \ac{TCGA} evaluation cohort (N=\nsamplestcgaeval{}; Methods~\ref{section:methods-tcga-cohort}), were validated and replicated as top-performing screening biomarkers, achieving a mean AUC of 0.88 (95\% CI: 0.87-0.90). Their performance in TCGA was comparable to that observed in MSK samples, with a mean AUC of 0.9 (95\% CI: 0.89-0.91; P-value = 0.41)
(Tables~\ref{table:results-merged},~\ref{table:msk-evaluation},~\ref{table:tcga-validation}).

We next evaluated OmniScreen's predictions of genomic alterations linked to therapeutic targets of \ac{FDA}-approved drugs across various cancer types, focusing on 54 treatment-associated genes with actionable hotspot mutations reported in \ac{MCG}~\cite{swanton2012my,holt2021my} and OncoKB~\cite{chakravarty2017oncokb,suehnholz2023cancer}
OmniScreen demonstrated strong predictive performance in identifying actionable targets that are routinely screened in clinical practice for therapy selection and response prediction (Figure~\ref{fig:S4}A, Table~\ref{table:results-hotspots}; Supplementary Note~\ref{section:supplementary-biomarker-therapy-targets}). For example, it accurately predicted BRAF V600E mutations, achieving AUCs of 0.93 and 0.89 in primary and metastatic thyroid cancers respectively, and performed well in primary cancers such as \ac{CRC} (AUC=0.91), glioma (AUC=0.88) and melanoma (AUC=0.76). Additionally, it effectively identified rare BRAF V600E carriers in NSCLC (N=11/903, 1.2\%; AUC=0.77). 
Other notable examples include the prediction of FGFR3 hotspot mutations (R248C, S249C, G370C, Y373C), which are associated with response to drugs such as erdafitinib, demonstrating high accuracy in bladder cancer (AUC=0.93 in primary samples and AUC=0.82 in metastatic samples).
It also predicted ERBB2 amplification, associated with response to trastuzumab, with an AUC of 0.81 in primary breast and metastatic gastric cancers. In metastatic \ac{NSCLC}, the model achieved an AUC of 0.79 and demonstrated 100\% \ac{NPV}, highlighting its potential to reliably identify patients unlikely to benefit from HER2-targeted therapy. 
In addition to ERBB2, OmniScreen showed robust performance in detecting \ac{TKI}-targetable hotspot mutations in primary \ac{NSCLC}, including EGFR (p.L858R, p.T790M, exon 19 deletions, and exon 20 insertions; AUC=0.87), KRAS p.G12C (AUC=0.83), MET exon 14 deletion/splicing mutations (AUC=0.79), and rarer RET fusion (N=12/903, 1.3\%; AUC=0.77). With an \ac{NPV} of 100\% for RET fusions, it enables the reliable exclusion of patients unlikely to harbor these alterations, supporting effective screening of both common and rare actionable mutations.

To evaluate phenotypes indicative of genomic instability, OmniScreen was examined on three key metrics: high tumor mutation burden (TMB-H), microsatellite instability or deficient mismatch repair (MSI-H/dMMR), and \acf{CIN} (Figure~\ref{fig:S4}C; Supplementary Note~\ref{section:supplementary-biomarker-genome-instability}). OmniScreen demonstrated strong performance in predicting TMB-H across nine cancer types (mean AUC = 0.88), with notable accuracy in primary \ac{CRC} and endometrial cancers (AUC = 0.90) and metastatic esophagogastric cancer (AUC = 0.87). Rare TMB-H cases in soft tissue sarcoma were also detected (N=8/346, 2.3\%; AUC = 0.89). The model demonstrated robust prediction of MSI-H/dMMR phenotypes across CRC (AUC = 0.97), endometrial (AUC = 0.87), esophagogastric (AUC = 0.96), and bladder cancers (AUC = 0.97), characterized by histopathological features such as poorly differentiated, medullary-like patterns of tumor growth with ``pushing'' borders, tumor infiltrating lymphocytes (TILs), and mucin production. OmniScreen also achieved high accuracy (AUC = 0.85 in primary \ac{CRC}) in identifying individuals with \ac{LS}, a hereditary condition linked to MSI-H/dMMR-associated cancers~\cite{elze2023microsatellite}. For CIN, a distinct form of genomic instability characterized by large-scale chromosomal alterations, OmniScreen showed a mean AUC of 0.84 in primary breast and ovarian cancers, with comparable performance in metastatic lesions.

\subsection{Tumor area and cohort size influence biomarker prediction performance}

\begin{figure}[hb!]
    \centering
    \includegraphics[width=\textwidth]{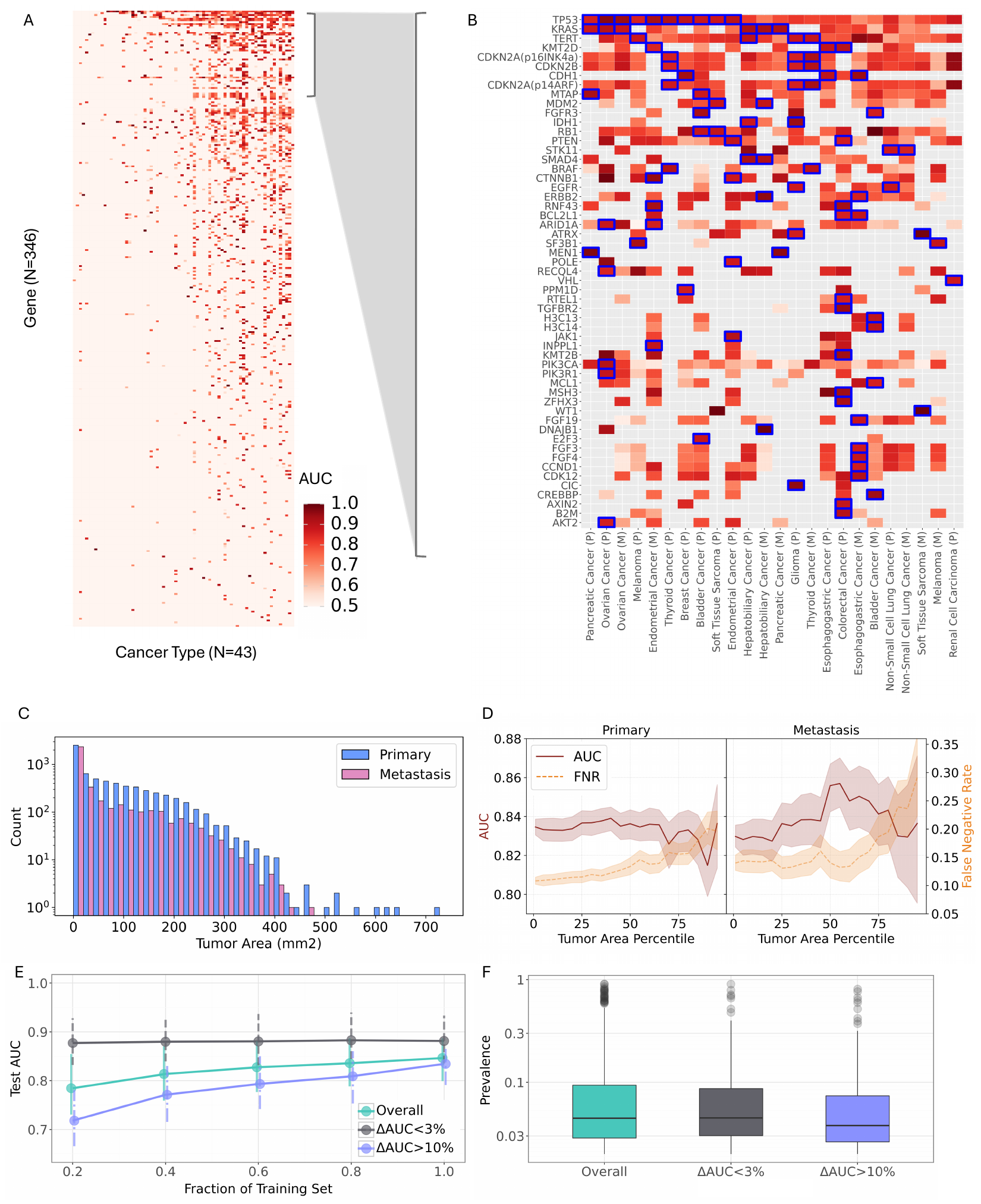}
    \caption{Pan-cancer digital biomarker prediction performance. \textbf{A}:~An overview heatmap showing the prediction of 739 genetic alteration biomarkers with AUC~$>0.5$, representing 320 genes across 43 cancers in the MSK test set. Biomarker labels included had more than 2 positive samples with a positive ratio~$>2\%$. Color coded by AUC. 
    \textbf{B}: Heatmap of gene biomarkers, with AUC~$>0.5$, identified from primary (P) and metastatic (M) cancers of the most common histologies. Blue boxes highlighted the top performing genes (N=47) in which the corresponding genetic alteration biomarkers (N=80) achieved an AUC~$>0.85$, sensitivity~$>0.8$ and specificity~$>0.3$.
    \textbf{C}: Distribution of tumor area in MSK test set, split by tumor site. 
    \textbf{D}: Change in AUC for biomarker prediction with respect to the increasing of tumor area within WSIs. 
    \textbf{E}: Change in AUC for biomarker prediction with respect to the fraction of training data set. 
    \textbf{F}: Distribution of biomarker prevalence (i.e., the ratio of positive samples) for high-performing biomarkers included in training data scale analysis.
    }
    \label{fig:2}
\end{figure}

To assess the impact of tumor area within \acp{WSI} (Figure~\ref{fig:2}C; Methods~\ref{method:tumorarea}) on biomarker predictions, we evaluated the AUC on samples with slides filtered by varying minimum tumor area thresholds. 
While certain biomarkers, such as MSI-H/dMMR (primary CRC), TMB-H (primary endometrial cancer), and DNAJB1 fusion (metastatic hepatobiliary cancer), exhibited performance invariant to tumor area filtering, the average AUC of the high-performing biomarkers showed a modest improvement after excluding slides with very small tumor areas ($\leq 10.05\,\text{mm}^2$, corresponding to the 35th percentile of tumor area in the tune set slides) (Figure~\ref{fig:2}D). 
However, this trend reversed, with a decline observed when slides with tumor areas $\leq 27.31\,\text{mm}^2$ (the 50th percentile in the tune set) were excluded, coinciding with an increase in the false negative rate.
False negative slides with tumor area $> 139.68\,\text{mm}^2$ (at 80th percentile) were found to be enriched for uncommon cancer types compared to slides with smaller tumor areas, based on pathologist's review.

In addition to tumor area, we evaluated the impact of training sample size on biomarker predictions by re-training our model on randomly sampled subsets of 20\%, 40\%, 60\% and 80\% of the training dataset. Model performance, assessed on the same test set, demonstrated an increase in AUC with larger training sample sizes (Figure~\ref{fig:2}E). For certain biomarkers, the test set AUC showed significant improvement ($\Delta\mathrm{AUC}>10\%$) with larger training sample sizes, for example, AKT1 (in metastatic prostate cancer, AUC increased from 0.43 to 0.77), TSC2 (in primary renal cell carcinoma, AUC increased from 0.52 to 0.98), and CTNNB1 (primary endometrial cancer, AUC increased from 0.77 to 0.87). This observation highlights the potential benefit of curating additional samples to further enhance their predictive performance. In contrast, biomarkers such as GNA11 mutation (primary melanoma), TP53 mutation (primary ovarian cancer), WT1 fusion (primary soft tissue sarcoma), MEN1 mutation (primary pancreatic cancer), and RB1 mutation (metastatic prostate cancer) exhibited relatively stable performance ($\Delta\mathrm{AUC}<3\%$) across varying training sample sizes. 
No difference was found in label prevalence between labels that showed minimal change in performance and those that exhibited a large increase in performance as the training sample size increased (Figure~\ref{fig:2}F), suggesting that the characteristics underlying the two groups of labels are independent of label prevalence.

\clearpage
\subsection{OmniScreen predictions are associated with histologic cancer subtypes}

\begin{figure}[!ht]
    \centering
    \includegraphics[width=\textwidth]{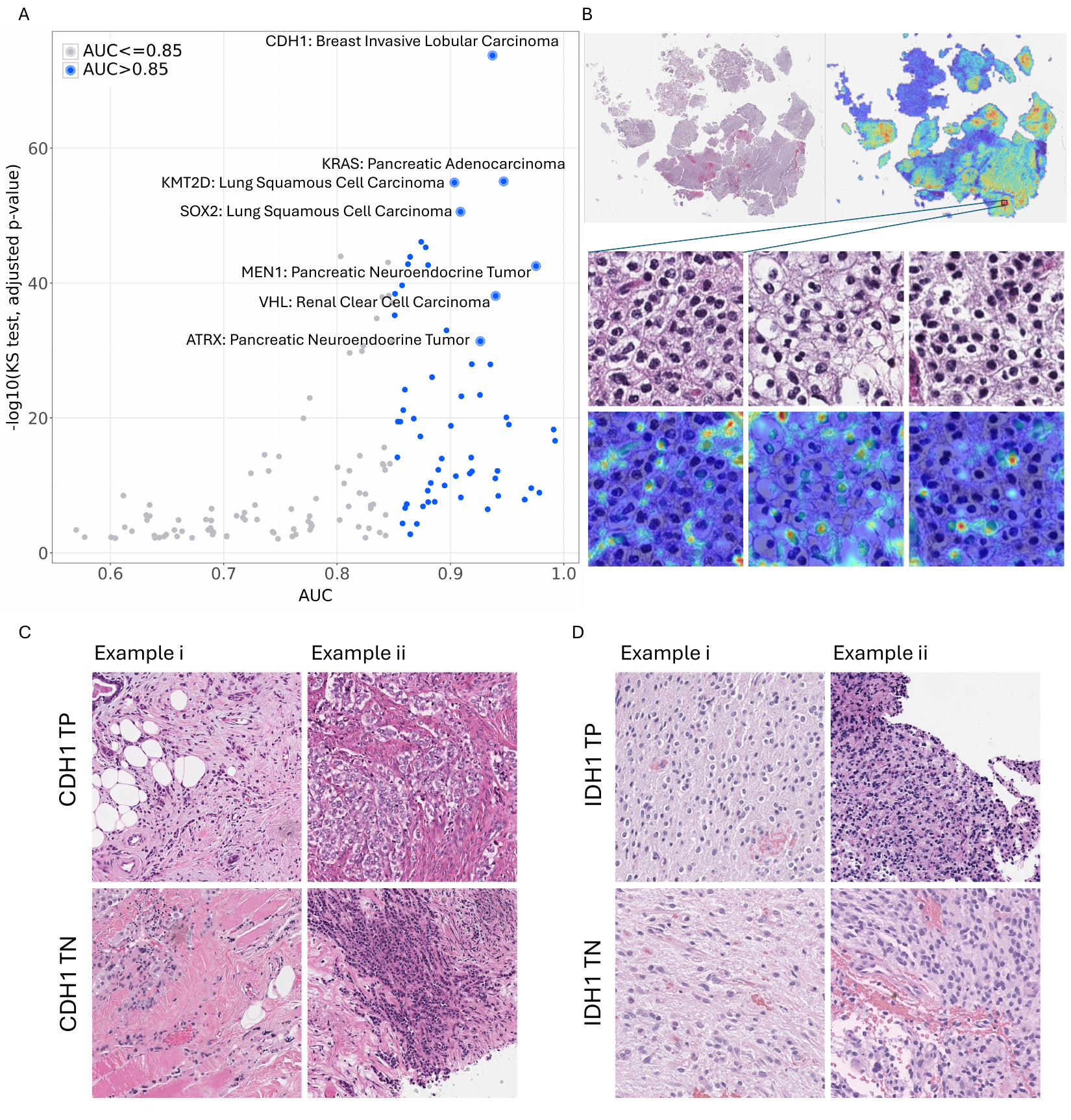}
    \caption{Biomarkers associated with histologic subtypes of cancers. 
    \textbf{A}:~$-log_{10}$(KS test, adjusted p-value) vs. AUC. Evaluation was performed using OmniScreen biomarker prediction scores and histologic subtypes as ground truth. Grey dots represent biomarkers with AUC~$\leq0.85$, blue dots denote those with AUC~$>0.85$. Text labels highlight example biomarkers stroongly associated with histologic subtypes, with $-log_{10}$(KS test, adjusted p-value)~$>30$ and AUC~$>0.9$.
    \textbf{B}: 
    Example of an \ac{HE}-stained WSI with an attention heatmap illustrating OmniScreen's prediction of IDH1 mutation likelihood for an oligodendroglioma, highlighting the targeted morphological features (regions highly attended by the model). The higher magnification tiles (top row) display the typical ``fried egg'' appearance of a classic oligodendroglioma. The attention heatmap (bottom row) highlights regions of high attention (red areas), focusing on the cytoplasmic clearing around the nucleus--characteristic of a formalin-fixed oligodendroglioma--and attention to at least one nucleus in each image. 
    Examples of $1024\times1024$ regions in WSIs that were highly attended by OmniScreen during the prediction of (\textbf{C}) CDH1 oncogenic mutations in breast cancer, and (\textbf{D}) IDH1 oncogenic mutations in glioma, including both true positive (TP) and true negative (TN) cases.
    \textbf{C}:~Representative images of CDH1-mutant cases (TP) showing the classic lobular invasive carcinoma histologic appearance with intracytoplasmic lumina and single file invasive carcinoma cells; and 
    CDH1-wildtype breast cancers (TN), exhibiting some of the single file and individual cell infiltration pattern.
    \textbf{D}:~Representative images of IDH1-mutant cases (TP) exhibiting classic features of oligodendroglioma, including (i) the classic cytoplasmic clearing of formalin fixed oligodendroglioma cells and (ii) less classic oligodendroglioma features; and IDH1-wildtype cases (TN), displaying morphological features characteristic of other glioma variants.
    }
    \label{fig:3}
\end{figure}

We next sought to determine whether OmniScreen could diagnose histologic cancer subtypes associated with the genomic alterations it was trained to detect. For each top-performing genomic biomarker within a cancer type, we compared the test set inference probabilities of the biomarker between the histologic subtypes. Our analysis revealed 52 subtype-specific biomarker predictions among 46 different genes (\ac{KS} test adjusted p-value~$<0.01$, and AUC~$>0.85$; Figures~\ref{fig:3}A-B, Table~\ref{table:results-histology}). We subsequently assessed the performance of these biomarkers in predicting the subtype diagnosis within the withheld test set.

OmniScreen accurately diagnosed subtypes by leveraging both classic histological features and subtle patterns less apparent to human pathologists.
For example, CDH1 alteration has a strong phenotype-genotype correlation with the invasive lobular carcinoma subtype of breast carcinoma (Figure~\ref{fig:S5}A). OmniScreen predicted the likelihood of CDH1 oncogenic mutations in breast cancers, accurately diagnosing invasive lobular carcinoma with an AUC of 0.93.
The regions highly attended to by OmniScreen when predicting CDH1 mutation in breast cancer exhibited classic histologic features of invasive lobular carcinoma, including intracytoplasmic lumina and single-file invasive carcinoma cells (Figure~\ref{fig:3}C). 
These features were necessary but insufficient to predict CDH1 mutation in breast cancer, as lobular carcinoma-like features such as single-file and individual cell infiltration patterns were also observed in CDH1-wildtype breast cancers. Despite this mimicry, OmniScreen accurately identified these cases as true negatives while also recognizing cases with less classic invasive lobular carcinoma features as true positives for CDH1. 
This association was further validated in the TCGA cohort, where it achieved an AUC of 0.95, and was thoroughly analyzed in one of our previous studies~\cite{pareja2024genomics}.

In glioma, oligodendroglioma is genetically defined by an IDH mutation and 1p19q codeletion~\cite{suwala2022oligosarcomas}. OmniScreen's prediction of IDH1 oncogenic mutation identified oligodendroglioma with an AUC of 0.89.
The attention heatmap (Figure~\ref{fig:3}B) reveals that OmniScreen focused on the cytoplasmic clearing around the nucleus, a hallmark of formalin-fixed oligodendrogliomas creating their classic ``fried egg'' appearance, and at least one nucleus per tile (highlighted in red). This suggests that the model associates high likelihood of IDH1 mutation with classic oligodendroglioma traits, consistent with the known high incidence of IDH1 alterations in this tumor type, while also capturing key subtle nuclear characteristics that may be less apparent to human pathologists. Notably, OmniScreen accurately predicted IDH1 mutation even in cases with less classic oligodendroglioma features (Figure~\ref{fig:3}D). As expected, \ac{TN} examples primarily featured other glioma variants without the distinct oligodendroglioma histology, which are less likely to harbor IDH1 mutations, thus enabling the model to reliably identify their IDH1-wildtype status.

Other notable genotype-phenotype associations identified by OmniScreen include VHL in \ac{ccRCC}, MEN1 and ATRX in pancreatic neuroendocrine tumors, KRAS in pancreatic adenocarcinoma, and GNAQ/GNA11 in uveal melanoma (see Supplementary Note~\ref{section:supplementary-biomarker-phenotype-association}). OmniScreen effectively detected a wide range of genomic alterations and histopathological features across both common and rare soft tissue sarcoma subtypes, such as RB1-loss in leiomyosarcoma, MDM2 amplification in liposarcoma, TERT in myxoid liposarcoma, and WT1 fusions in desmoplastic small round cell tumors (DSRCT). Furthermore, it distinguished molecular subsets within histological subtypes, such as SOX2 amplification and KMT2D alterations in \ac{LUSC}, and identified shared pathological features, such as ARID1A deleterious mutations across \ac{OCCC} and endometrioid ovarian cancers.

\subsection{OmniScreen captures phenotypes linked to shared signaling pathways}

\begin{figure}[hb!]
    \centering
    \includegraphics[width=\textwidth]{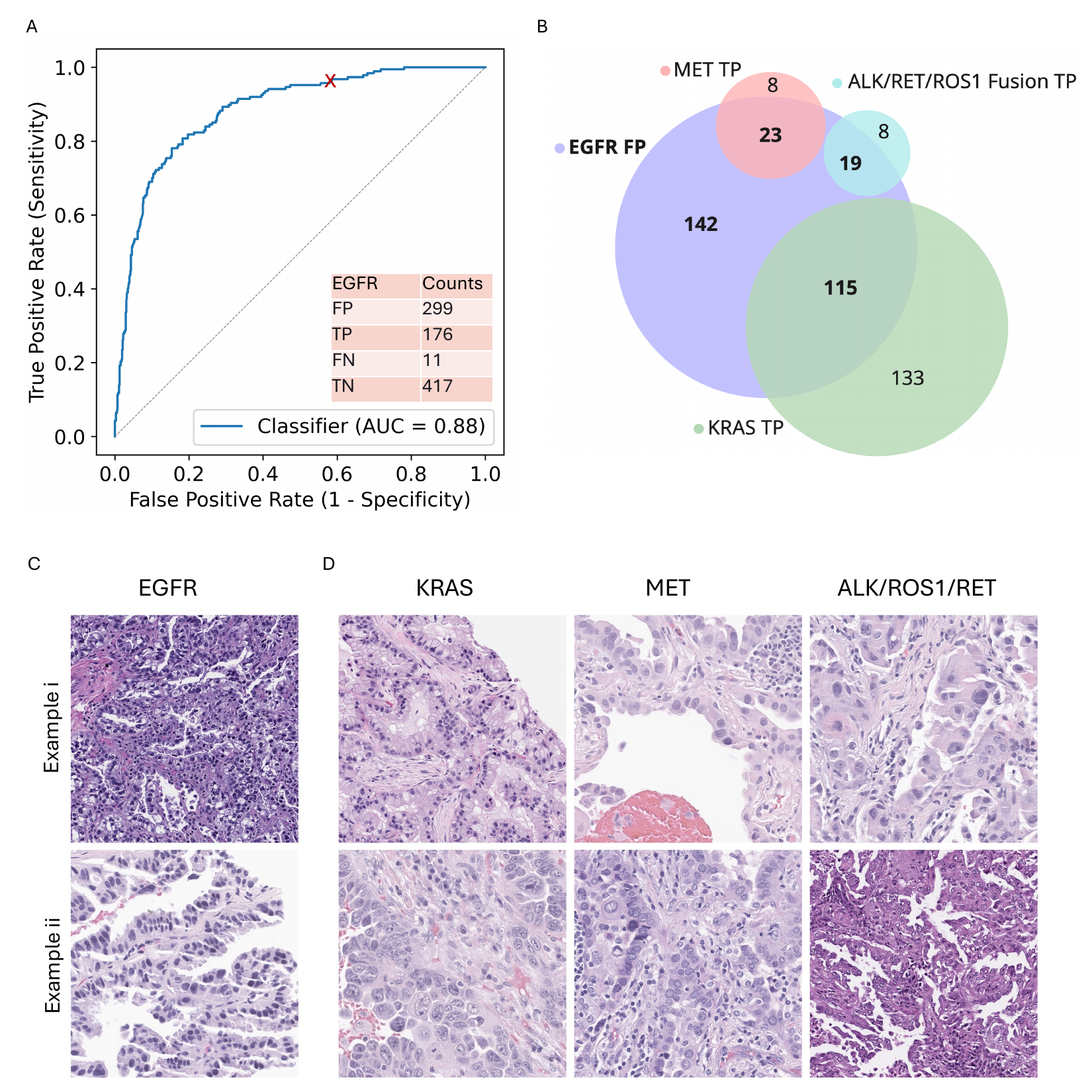}
    \caption{
    NSCLC EGFR panel prediction. 
    \textbf{A}:~ROC curve of EGFR oncogenic mutation prediction. $\times$ indicates the operating point at 90\% sensitivity. Classification of binary predictions: true positives (TP), true negatives (TN), false positives (FP), false negatives (FN) were determined against ground truth confirmed by MSK-IMPACT sequencing.
    \textbf{B}:~Euler diagram illustrating the the overlap of EGFR FP samples that are TP for other genes within the same pathways as EGFR in NSCLC, including KRAS (N=115), MET (N=23) and ALK/RET/ROS1 fusion (N=19). 
    Examples of $1024\times1024$ regions in lung adenocarcinomas slides that were highly attended by OmniScreen during the prediction of EGFR oncogenic mutations, 
    \textbf{C}:~EGFR TPs, showing histologic patterns consistent with a likelihood of EGFR alteration, such as lepidic (ii), acinar (i), and papillary/micropapillary growth patterns were evident, as were moderate (i, ii) to high nuclear atypia;
    \textbf{D}:~EGFR FPs, predicted with high EGFR mutation likelihood by OmniScreen, confirmed as EGFR wild type and harboring KRAS, MET, or ALK/ROS1/RET mutations. These cases predominantly exhibit acinar (KRAS: i, ii; MET: ii), papillary (MET: i; ALK/ROS/RET: ii), solid-type growth (ALK/ROS/RET: i) patterns, with moderate or greater (KRAS: ii; ALK/ROS/RET: i) nuclear pleomorphism and atypia. 
    }
    \label{fig:4}
\end{figure}

In NSCLC, alterations in EGFR, BRAF, RAS, MET, ALK, ROS1, and RET signal through a shared molecular pathway associated with activated receptor tyrosine kinase activity. This leads to activation of downstream signaling pathways such as MAPK/ERK and parallel signaling via PI3K/AKT. We investigated whether mutations in genes that share a signaling pathway could result in a common phenotype detectable by OmniScreen. When using an operating point optimized for 90\% sensitivity, 475 EGFR positive predictions were identified in NSCLC in our test set (Figure~\ref{fig:4}A). Among these positive predictions, 176 were confirmed as EGFR-mutant cases by MSK-IMPACT sequencing. However, the remaining 299 EGFR cases were instead classified as EGFR wild-type by MSK-IMPACT. Among these EGFR \acp{FP}, 157 harbored alterations in other genes commonly mutated in NSCLC that signal through a shared molecular pathway with activating EGFR mutations (N=115 KRAS oncogenic mutations, N=23 activating MET and N=19 ALK/RET/ROS1 fusions; Figure~\ref{fig:4}B).
These results are consistent with an AI model detecting a phenotype associated with the net phenotypic effect of activating oncogene mutations via their shared downstream signaling activity rather than distinguishing individual gene-specific alterations in members of the shared signaling pathway. 

While reviewing the corresponding WSIs, we observed that the regions highly attended to by OmniScreen for \ac{TP} EGFR prediction in NSCLC predominantly contained adenocarcinomas, as expected, and exhibited lepidic, acinar, and papillary/micropapillary growth patterns, along with moderate to high nuclear atypia consistent with features associated with \ac{TP} EGFR alterations (Figure~\ref{fig:4}C). 
In the EGFR \ac{FP} slides harboring KRAS mutations, MET mutations, or ALK/RET/ROS1 fusions, the predominant morphologic features included acinar and solid-type growth patterns, along with moderate to marked nuclear pleomorphism and atypia (Figure~\ref{fig:4}D). Notably, these histologic characteristics overlap substantially with those seen in EGFR-mutated lung adenocarcinomas, potentially reflecting the shared molecular signaling cascades associated with these genetic alterations. 

Building on our EGFR hypothesis in NSCLC, we hypothesized that mutations in any member of molecular pathways may result in overlapping, shared phenotypes due to the interconnected nature of their signaling cascade. Supporting this hypothesis, OmniScreen successfully predicted, per case, the presence of oncogenic genomic alterations in any of the genes within canonical pathways
(Figure~\ref{fig:S4}C, Tables~\ref{table:msk-pathways},~\ref{table:msk-evaluation}).
For example, RTK/MEK/ERK pathway alterations were predicted with AUC~$>0.75$ in primary samples of glioma, \ac{NSCLC}, prostate and breast cancers. Similarly, mTOR pathway alterations were detected in primary endometrial, thyroid, and \ac{RCC} (AUC~$>0.75$). In primary \ac{CRC}, OmniScreen identified alterations in \ac{HRD}, TGF-$\beta$, and \ac{DDR} with a mean AUC of 0.80 (range: 0.78-0.82).

\subsection{OmniScreen improves prediction of clinically relevant low-prevalence biomarkers}

\begin{figure}[ht]
    \centering
    \includegraphics[width=\textwidth]{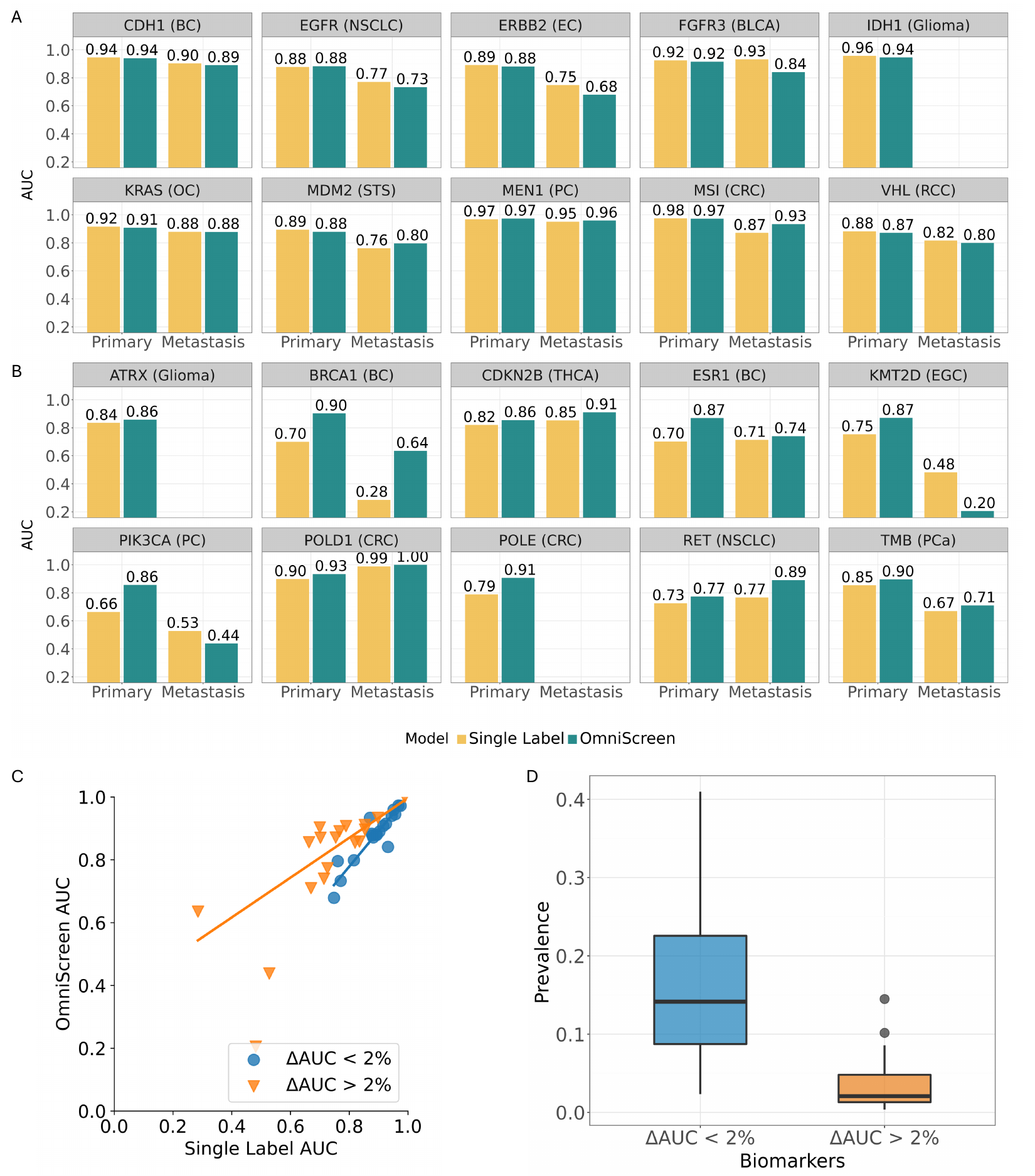}
    \caption{Comparing biomarker prediction performance (AUC) achieved by different models: conventional individually trained models (Single Label) within specific tumor types vs. OmniScreen. 
    Selected biomarkers stratified into two groups: \textbf{A}:~biomarkers (N=10) with $\Delta\mathrm{AUC}$ (relative change in AUC between two models)~$<2\%$, and \textbf{B}:~biomarkers (N=10) with $\Delta\mathrm{AUC}>2\%$. Tumor types including:
    Breast Cancer (BC), Bladder Cancer (BLCA), Colorectal Cancer (CRC), Endometrial Cancer (EC), Esophagogastric Cancer (EGC), Non-Small Cell Lung Cancer (NSCLC), Ovarian Cancer (OC), Pancreatic Cancer (PC), Prostate Cancer (PCa), Renal Cell Carcinoma (RCC), Soft Tissue Sarcoma (STS), Thyroid Cancer (THCA). 
    \textbf{C}: Correlation between AUC values obtained from the single-label models and OmniScreen. Fitted lines are shown for each group of biomarkers to highlight trends within the data.
    \textbf{D}: Comparison of biomarker prevalence distributions between two groups: biomarkers with $\Delta\mathrm{AUC}<$2\% and biomarkers with $\Delta$AUC$>2$\%. 
    }
    \label{fig:5}
\end{figure}

To compare the performance of OmniScreen with conventional approaches which train individual models for each biomarker within a specific cancer type, we selected 20 biomarkers spanning multiple tumor types, representing a broad range of prevalence and trained a single model for each (Methods~\ref{method:singlelabel}, Figures~\ref{fig:5}A-B).
 
Overall, the AUC values of biomarkers generated by OmniScreen correlated strongly with those obtained using the conventional method (Figure~\ref{fig:5}C). 
However, notable variation was observed in the relative change in AUC (i.e., $\Delta\mathrm{AUC}$) across biomarkers. Stratifying biomarkers into two groups based on $\Delta\mathrm{AUC}$~($<2\%$ and~$>2\%$) revealed that biomarkers with smaller relative changes in performance were associated with higher prevalence, whereas those with larger relative changes corresponded to lower prevalence (Figure~\ref{fig:5}D). 
These findings suggest that for high-prevalence biomarkers, the conventional approach yielded AUC values comparable to those achieved by OmniScreen, while OmniScreen demonstrated superior average performance for low-prevalence biomarkers, albeit with slightly greater variability.


\section{Discussion}


Genomic alterations can give rise to distinct cancer phenotypes, some of which manifest as specific morphological features visible under light microscopy. For example, bi-allelic inactivation of E-cadherin (CDH1) in breast cancer produces a characteristic discohesive single-file pattern, recognized by pathologists as invasive lobular carcinoma~\cite{dopeso2024genomic,pareja2024genomics}. However, certain genomic alterations may result in phenotypes that are either too subtle, or too overlapping, to be reliably detected by routine human assessment. 

The integration of machine learning into digital pathology offers a transformative solution, enabling scalable analysis of WSIs to uncover subtle phenotype-genotype correlations. Leveraging large, expertly annotated datasets and advanced computational methods, AI-driven models can identify digital biomarkers associated with clinically relevant genomic alterations. This enhances diagnostic precision, informs therapeutic decision-making, and accelerates clinical research by generating novel phenotype-genotype hypotheses and refining patient selection for clinical trials~\cite{lyman2016biomarker,zhou2024tumor}.

In this study, we introduce OmniScreen, a robustly performing \ac{AI}-based model trained on histological and molecular data from \ac{MSKCC} and \ac{TCGA}. 


OmniScreen was successfully evaluated on the 15 most common cancer types treated at MSKCC, using molecular ground truth from the MSK-IMPACT \ac{NGS} assay. It demonstrated strong performance in detecting clinically relevant genomic biomarkers routinely screened for targeted therapy selection and response prediction, including hotspot mutations in \ac{NSCLC} (EGFR, KRAS, MET, and RET), ERBB2 amplification in breast and gastric cancers for HER2-targeted therapy, FGFR3 alterations in urothelial carcinoma, and BRAF V600E mutations in thyroid cancers, melanoma and \ac{CRC}. 

The model demonstrated robust performance in predicting MSI-H/dMMR phenotypes across colorectal, endometrial, gastric, and bladder cancers, as well as detecting \ac{TMB-H} in colorectal and endometrial cancers. It identified rare \ac{TMB-H} cases in low-prevalence histologies, such as soft tissue sarcoma~\cite{shao2020prevalence}, highlighting its potential to identify patients who may benefit from \ac{ICI} therapy, even in less frequently screened tumor types. These findings support the use of AI-assisted digital biomarker screening with \ac{HE}-stained WSIs as a cost-effective, time-efficient, and tissue-sparing approach to prioritize patients for definitive molecular testing. 

While certain genes did not reach high AUCs, when applying a threshold for high-sensitivity screening (sensitivity $>0.9$), predictions for genomic alterations, such as RET fusions and ERBB2 amplifications in NSCLC, demonstrated a high NPV of 100\%. 
This suggests their potential as ``rule-out'' digital biomarkers, for triaging patients unlikely to harbor these mutations to reduce unnecessary molecular testing.

As proof of concept, we estimated cost savings for enrolling 500 patients, pre-screened by OmniScreen, with all confirmed positive via definitive NGS or PCR (see Supplementary Note~\ref{supplementary-method:costanalysis}). The AI-assisted triage demonstrated potential average savings of \$320K (5-30\% reduction) for PCR and \$1,086K (13-39\% reduction) for NGS (Figure~\ref{fig:S4}B). Cost savings were most pronounced for low-prevalence targets, highlighting the potential of AI-assisted triage to overcome financial barriers in clinical studies targeting rare mutations.
Beyond clinical trials, the digital biomarker screening approach offers opportunities to reduce routine testing costs in specific laboratory and practice settings by optimizing the use of more expensive definitive tests.

Our results demonstrate the expected bias of genomic alterations associated with specific phenotypes and histologic subtypes, such as MDM2 amplification in liposarcoma, VHL loss in clear cell renal carcinoma, and IDH1 prediction in oligodendroglioma.
These results confirm that the model reliably identifies established diagnostic phenotype-genotype correlations, effectively training itself on known phenotypes. The consistent detection of expected biases across histologies further supports the validity of the model's predictions, even for genomic alterations without previously described phenotypic associations.

Variations in the accuracy of biomarker predictions across tumor types may be attributed to differences in sample size, mutation prevalence, and the intrinsic strength of phenotypic manifestations associated with specific biomarkers. 

Our development cohort included a higher proportion of primary samples compared to metastatic samples, which may account for the generally better performance observed in primary samples across the tested histologies. Expanding the dataset to include more samples from metastatic lesions may uncover additional biomarkers associated with metastasis, and potentially enhance the accuracy of genomic alteration predictions in metastatic tumors.

While some genomic alterations were represented by relatively small numbers of cases in our dataset, biomarkers with strong genotype-phenotype associations, such as GNA11 in melanoma, MEN1 in pancreatic cancer, and WT1 fusion in soft tissue sarcoma, exhibited distinct morphological features that were reliably detectable even with limited sample sizes. 
However, not all genomic alterations exhibit distinct or robust phenotypes, which can constrain their detectability. For such biomarkers, increasing the number of cases with these alterations during model training could amplify the signal detection and further improve prediction performance.



As low tumor content may influence the accuracy of biomarker prediction, we investigate its impact by using tumor area as a surrogate marker for tumor content. While the AI model may struggle with slides containing insufficient tumor area, an exceptionally large tumor area did not consistently improve performance. Some biomarkers, such as AR amplification (metastatic prostate cancer), CCND1 amplification (primary gastric cancer), and BRAF oncogenic mutation (primary melanoma), were more sensitive to limited tumor area than others. 
This suggests that biology, in terms of the robustness of a genomic alteration associated phenotype, is a significant, if not controlling, factor in ability to train AI to predict features such as underlying genomic alterations. 
A strong, robust phenotype associated with a genomic alteration may be identifiable via AI on WSIs even with small tumor areas. Conversely, a genomic alteration with a weak phenotypic effect on the tumor in \ac{HE} WSIs may require a greater minimum amount of tumor present for accurate identification. If the phenotype is too weak or lacks sufficient differentiation, even large tumor areas may not improve performance. Some biomarkers may not have a robust enough phenotype to ever be learnable by ML on WSIs to a level that would allow clinical use. 

Furthermore, we found in some cases that slides with extremely large tumor areas sometimes skewed training or were underrepresented in the training set but appeared in the test set. In rare tumor types, small or large biopsies may be over- or underrepresented due to limited cases for training. Ensuring AI models are trained on a diverse range of tumor phenotypes, including variations in tumor amounts, sample sizes, and primary versus metastatic settings, is critical for capturing phenotype variability.

Unexplored factors such as crush/cautery artifacts or prolonged tissue ischemia may impact the phenotype on \ac{HE} WSIs. This underscores the need to consider phenotypic and spatial diversity, not only tumor size, when evaluating biomarker performance.

The robustness and clinical utility of AI applied to WSIs may also be restricted by biology in other ways. For example, where common gene activating alterations converge on a shared oncogenic molecular pathway, such as the commonly therapeutically targeted genes in NSCLC (EGFR, KRAS, MET, BRAF, ERBB2 and ALK/ROS1/RET), distinguishing gene-specific phenotypes based solely on morphologic features can be challenging, even for AI. The phenotype may be most robust for the shared molecular endpoint, rather than the specific gene of origin, making an AI algorithm best at identifying if any of the convergent genes are altered versus discriminating which gene was altered to create the phenotype. However, this highlights an opportunity to leverage AI-assisted methods to screen panels of related genes in a single test via activation of the shared molecular pathway by identifying the `pathway-active' phenotype. 
When combined with definitive molecular testing, it could prioritize cases for confirmatory sequencing of panel genes to identify specific gene alterations, as well as identify mutations functionally linked to pathway activation, enhancing diagnostic precision and improving therapy response prediction, beyond genotyping alone. 

In summary, we describe a fully automated method for rapid high-throughput AI-assisted screening of cancer WSIs, not only to detect clinically meaningful genomic abnormalities across histologies, but also to identify the histologies characterized by genomic alterations. This may select patients for definitive genomic analysis and/or enable phenotype-genotype correlations to improve clinical decisions. This pipeline is cost and time efficient, tissue sparing and capable of application to large patient populations across a wide range of cancer with reliable performance metrics. Such an approach is suitable for application in research, including clinical trial settings.

\section{Methods}
\label{section:methods}

\subsection{Patient cohorts, histopathologic and genomic analyses}
\label{section:methods-cohorts}

\subsubsection{MSK cohort: MSK IMPACT Dataset}
\label{section:methods-msk-cohort}

A sample is defined as an MSK-IMPACT assay paired with one or more \ac{HE} stained WSIs taken from the same formalin-fixed paraffin-embedded (FFPE) tissue block. The MSK-IMPACT dataset consists of \nsamples{}~samples (\nslides{}~\acp{WSI}) from 72 cancer groups (71 cancer types and an ``Unknown'' category for which the cancer type is not known). \nprimary{} (65\%) of samples are primary cancer sites while the remaining \nmetastasis{} (35\%) are metastasis samples. All image data were retrieved from the hospital archives and verified to meet staining quality standards for histopathology review.
All biopsy glass slides were scanned with Leica Aperio AT2 scanners (Leica Biosystems, Division of Leica Microsystems Inc, Buffalo Grove, IL, USA) at 20$\times$ (0.5 microns per pixel) or 40$\times$ (0.25 microns per pixel) magnification. See Table~\ref{table:msk-counts} for a detailed breakdown of patient, sample and slide counts by cancer type and tumor site in the dataset.


The cohort underwent paired tumor-normal targeted sequencing using the \ac{FDA}-cleared \ac{MSK-IMPACT} assay~\cite{cheng2015memorial}. MSK in-house bioinformatic pipelines were employed to analyze the sequencing results. The analysis determined the genomic alteration status, including mutations (\ac{SNVs} and \ac{indels}), copy number variations (amplification and deletions), and structural rearrangements (fusions) in \nimpactgenes{} important cancer associated genes. 
The list of genes was compiled from four MSK-IMPACT versions including v3 (N=341 genes), v5 (N=410), v6 (N=468 genes) and v7 (N=505 genes).
Additional fusion events confirmed by the MSK-Fusion panel~\cite{hechtman2017pan,zhu2019diagnosis} via RNA sequencing were incorporated to improve the coverage of fusion detection for the genes that are available from MSK-IMPACT panel. The annotation of the oncogenicity and clinical relevance of specific genetic alterations was determined using OncoKB~\cite{chakravarty2017oncokb,suehnholz2023cancer}. \Ac{TMB} score, \ac{MSI} score, \ac{LOH}, and \ac{WGD}, where available, were analyzed and retrieved at \ac{MSK}. 
Furthermore, the status of \ac{ER}, \ac{PR}, \acp{HER2}, \ac{PD-L1}, and \ac{dMMR} were confirmed by \ac{IHC}.

Among \ac{MSK-IMPACT} panel genes, five groups of genes that canonically participate in shared pathways associated with DNA repair mechanisms were identified. The gene members of each pathway were retrieved from \ac{MCG}~\cite{swanton2012my,holt2021my}. Table~\ref{table:msk-pathways} summarizes the list of gene candidates associated with each pathway, including DNA damage response (DDR), receptor tyrosine kinase (RTK) MEK/ERK pathway, homologous recombination deficiency (HRD) pathway, mTOR signaling pathway, and TGF-$\beta$ signaling pathway. 



\subsubsection{TCGA cohort: TCGA PanCancer Atlas}
\label{section:methods-tcga-cohort}

The external cohort consists of samples from \ac{TCGA} PanCancer Atlas project~\cite{weinstein2013cancer}. A total of \nslidestcgaoriginal{} diagnostic \acp{WSI}, corresponding to 32 cancer types, were retrieved from \ac{TCGA}. The corresponding genomic alteration data, including mutation, copy number aberration, and fusion, were downloaded from cBioPortal (\url{http://www.cbioportal.org}).
A total of 400 slides were excluded from model training and evaluation. This included 360 slides that have no associations with any biomarkers, 36 slides missing \ac{mpp} information, and 4 slides that were out of focus and had no foreground tiles detected. Of the remaining samples, \nsamplestraintcga{}  (\nslidestraintcga{} \acp{WSI}) were used in training and \nsamplestcgaeval{} (\nslidestcgaeval{}~\acp{WSI}) define the validation cohort. See Table~\ref{table:tcga-counts} for a detailed breakdown of patient, sample and slide counts by cancer type in the dataset.

\subsection{Development of an AI-based system for the detection of pan-cancer digital biomarkers using whole slide images}
\label{method:model}

We developed OmniScreen, a pan-cancer digital biomarker screening model to predict genomic abnormalities of interest in human cancers from \ac{HE} \acp{WSI}. 
The model was trained on \nslidestrain{} diagnostic clinical \acp{WSI} from a cohort of \npatientstrain{} patients (N=\npatientstrainmsk{} treated at \ac{MSK}; N=\npatientstraintcga{} \ac{TCGA} patients). The training cohort covers 71 different cancer types with \nimpactgenes{} genes assessed by the MSK-IMPACT targeted sequencing oncology assay (Figure~\ref{fig:S1}, Table~\ref{table:dataset-counts}). 
\nvectorlabels{} training ground truth labels include oncogenic point mutations, copy number variations (amplifications or deletions) and fusion events, or the presence of these types of genetic variations in any of a group of genes canonically participating in a shared signaling pathway associated with cancer, e.g., DNA damage responses, \ac{RTK} pathway, and mTOR signaling pathway (Figure~\ref{fig:S2}, Table~\ref{table:msk-pathways}).

The task is framed as multi-label binary classification, where genomic features (biomarkers) are represented as binary labels. Each label indicates the presence or absence of genomic alterations in a single gene, or in any of a group genes that participate in a shared signaling pathway. The genomic feature binary labels derived from MSK-IMPACT results are oncogenic mutations, copy number amplification, copy number deletions, fusions, or the combination of oncogenic mutation/amplification if a gene is an oncogene and oncogenic mutation/deletion if a gene is a \ac{TSG} (Figure~\ref{fig:S2}). Additional genomic features included in training are \ac{TMB-H} for \ac{TMB}\,$\geq10$ mutations/megabase (mut/Mb), \ac{MSI-H} for \ac{MSI} score\,$\geq10$, and \ac{MSS} for \ac{MSI} score\,$<3$, \ac{dMMR} for loss of IHC staining in \ac{MMR} (MLH1, MSH2, MSH6, and PMS2) proteins and harbored genetic alterations in \ac{MMR} genes, Lynch Syndrome for the presence of germline mutation in any of MLH1, MSH2, MSH6, PMS2, EPCAM, 
and \ac{CIN} measures: tetraploidy, whole genome doubling (WGD), loss-of-heterozygosity (LOH) in~$\geq50\%$ genome, fraction of genome altered (FGA)~$\geq30\%$, and genome instability index (GI index)~$\geq20\%$. The GI index is a metric derived from \ac{FGA} and \ac{LOH}, ranging from 0 to 1. 

The MSK development cohort was split into train, tune and test datasets with a ratio of 4:1:1. This partitioning resulted in sample sizes of \nsamplestrain{} (\nslidestrain{}~WSIs) for the train set, \nsamplestune{} (\nslidestune{} WSIs) for the tune set, and \nsamplestest{} (\nslidestest{}~WSIs) for the test set. 
The distribution of cancer types across the datasets included 71 different types in the train set, 63 in the tune set, and 62 in the test set (Figures~\ref{fig:1}B,~\ref{fig:S1}B, Table~\ref{table:msk-counts}). 
Each dataset also includes the category of an ``Unknown'' histology. A total of \nvectorlabels{} biomarker labels with at least 8 positive and 8 negative samples in the train set and at least 4 positive and 4 negative samples in the tune set were included in the training. This corresponds to \nimpactgenes{} genes out of 505 MSK-IMPACT panel genes.
To enhance the model's generalizability, the TCGA cohort was divided into training and evaluation sets. Specifically, 80\% of the TCGA cases were combined with the MSK train dataset to form the training cohort, while the remaining 20\% of TCGA cases served as an independent validation set.

Each slide is split into image tiles of size $224\times224$ pixels. The tiles are filtered to only include those representing foreground (tissue) using a foreground detection model based on a \ac{FCN}. Each foreground tile is embedded with the foundation model Virchow2~\cite{vorontsov2024foundation,zimmermann2024virchow2scalingselfsupervised} into a tile embedding of length 1280. Each slide is thus represented as a $N\times1280$ tensor, where $N$ is the number of foreground tiles.

The embeddings serve as input into a feed-forward network with an attention mechanism that aggregates the tile-level embeddings into a slide-level prediction~\cite{vorontsov2024foundation}.
See Supplementary Note~\ref{section:supplementary-methods-omniscreen} for an overview of OmniScreen model and training parameters. The model was trained on slide level. The final prediction per sample was determined as the maximum prediction over slides in the sample.

Checkpoint selection was done on the tune set using the mean AUC and mean \ac{AP} across labels that contained at least 5 positive samples. The operating threshold for each label was determined on the tune set by optimizing for 90\% sensitivity in each cancer type present in the tune set. A threshold thus corresponds to a label and cancer type pair. These thresholds were then applied to generate sample-level binary predictions from the inference probabilities in the MSK test set and the TCGA validation set, indicating the presence or absence of the genetic mutation.


\subsection{Tumor area estimation in WSIs using the Paige PanCancer Detect}
\label{method:tumorarea}
We applied Paige PanCancer Detect~\cite{vorontsov2024foundation} to generate cancer and precursor tissuemaps for each slide in the tune and test sets. 

The Paige PanCancer Detect model consists of a feature embedder (Virchow2~\cite{zimmermann2024virchow2scalingselfsupervised} was used) and a tile classifier network. The tile classifier is a small fully connected network trained on embeddings of 1.4M slides containing 54 tissue types, using a combination of strong and weak supervision. In order to overcome different levels of data imbalance across different tissue types, multiple networks were trained on diverse subsets of tissue types. 

During inference, an ensemble of 8 linear models is used to classify cancer and an ensemble of 4 models is used to classify precursor lesions. Each slide is divided into $224\times224$ tiles, and tile embeddings are generated using Virchow2. These embeddings are processed by both ensembles to compute cancer and precursor scores for each tile. The scores are then arranged based on the spatial coordinates of the tiles to produce probability maps that represent their spatial distribution. A threshold of 0.1 is applied to the probability maps, followed by smoothing to convert them into tissuemaps--overlay images highlighting all detected cancer or precursor regions on the slide. 

The tumor area for a WSI was estimated as the union of cancer and precursor regions. Tumor area thresholds were determined based on the 1st, 5th, 10th, and up to the 95th percentiles of tumor areas in the slides from the tune set. These thresholds were then applied to filter slides in the test set, enabling the evaluation of biomarker performance on a subset of samples.

\subsection{Training single biomarker label models}
\label{method:singlelabel}
To evaluate the performance of OmniScreen relative to the conventional approach—where models are trained separately for individual biomarker labels within specific cancer types—we selected a panel of 20 biomarkers that are clinically relevant or well-established in the literature for specific tumor types. For each biomarker, we trained a model on a subset of the training dataset restricted to the associated tumor type. 20 single label biomarker models were then trained. The models were evaluated using the relevant subsets of the tune and test datasets.

See Supplementary Note~\ref{section:supplementary-methods-single-label} for an overview of the model and training parameters. The models were trained on slide level. The final sample-level prediction was determined as the maximum prediction over slides in the sample.

When comparing the conventional approach with OmniScreen, we computed $\Delta$AUC, i.e., the relative change in AUC, to quantify the difference in performance between two approaches.
\[\Delta~AUC = \left|\frac{AUC_{OmniScreen} - AUC_{SingleLabel}}{AUC_{SingleLabel}}\right|\]
Due to the limited availability of metastatic samples for certain biomarkers, we ranked the biomarkers based on their $\Delta\mathrm{AUC}$ in primary samples. Using the median $\Delta\mathrm{AUC}$ value of 2\%, we divided the biomarkers into two groups: $\Delta\mathrm{AUC}<2\%$ and $\Delta\mathrm{AUC}>2\%$.

\subsection{Phenotype-genotype correlation analysis}
\label{method:genophenocorrelation}
For each histology subtype in a cancer, we used the Kolmogorov–Smirnov (KS) one-side test to examine whether the inference probabilities of a biomarker label in the target subtype is greater than the inference probabilities in the other subtypes of cancer. The p-values were adjusted using the Benjamini-Hochberg method. P-values\,$<0.01$ were considered as statistically significant. A histologic subtype ground truth is defined as a binary label indicating if the sample was annotated for a given histologic subtype. For a biomarker with significantly higher inference probabilities in a specific histologic subtype of cancer, the AUC of the biomarker in prediction of the corresponding subtype was evaluated by using inference probabilities as prediction and the binary label of histologic subtype as ground truth. The sensitivity and specificity was computed by using the binary prediction of the genomic alterations as prediction label and the binary label of histologic subtype as ground truth.

\clearpage

\section{Acknowledgement}
We would like to extend our sincere gratitude to several individuals who contributed significantly to the success of this project. Philippe Mathieu, Alexander van Eck, Wayne Hendricks, Sid Senthilnathan, Michael Singer and Eric Robert played a crucial role in developing the AI and data infrastructure, which enabled the large-scale training and evaluation essential to this work.

We also wish to acknowledge Jan Bernhard for his earlier contributions to the development of digital biomarker projects related to MSI detection, which laid the foundational work for this project.

Our thanks also go to Eugene Vorontsov, Julian Viret, Adam Casson, George Shaikovski and Donghun Lee for their invaluable work in the development of the foundation model Virchow2 and PanCancer Detect.

We would like to thank Dani Gorton for her valuable advice, as well as her time and effort in reviewing the design and editing of Figure~\ref*{fig:1} included in this study.

Finally, we are deeply grateful to Brandon Rothrock for his supervision and mentorship during the early stages of the biomarker-related research, which significantly influenced the direction of this research.

\section{Data protection}
This project was governed by an Institutional Review Board approved retrospective research protocol, under which consent/authorization was waived before research was carried out. Data collection was conducted exclusively at MSK. The AI model was developed by Paige. 


\clearpage
\section{Tables}

\begin{table}[ht]
\caption{Development and validation datasets - sample and slide counts}
\label{table:dataset-counts}
\centering
\begin{tabular}{lrrrrrrr}
\toprule
 & Train & & Tune & & Test & \\
\midrule
Dataset & Samples & WSIs & Samples & WSIs & Samples & WSIs \\
\midrule
MSK  & \nsamplestrainmsk  & \nslidestrainmsk  & \nsamplestune & \nslidestune & \nsamplestest & \nslidestest \\
TCGA & \nsamplestraintcga & \nslidestraintcga &             - &      -       & \nsamplestcgaeval & \nslidestcgaeval \\
\bottomrule
\end{tabular}
\end{table}




\clearpage
\section{Supplementary Information}
\setcounter{table}{0}
\renewcommand{\thetable}{S\arabic{table}}

\subsection{Supplementary Methods}
\label{section:supplementary-methods}
\subsubsection{OmniScreen Aggregator Training}
\label{section:supplementary-methods-omniscreen}
The aggregator model code can be accessed at \url{https://github.com/Paige-AI/paige-ml-sdk}.
The aggregator model was trained with model parameters described in Table~\ref{table:model-params}. The model was trained for 50 epochs with training parameters described in Table~\ref{table:training-params}. The input embedding (generated by Virchow2) consists of class token only (dimension 1280). The training took approximately 52 hours and was performed on 2 NVIDIA A100 (40GB) nodes (8 GPUs per node).
\begin{table}[ht]
\caption{Aggregator model parameters}
\label{table:model-params}
\centering
\begin{tabular}{l|l|r}
     \toprule
     Parameter description & Parameter name & Parameter value \\
     \midrule
     Dimension of the input embedding & \texttt{in\_features} & 1280 \\
     Dimension of first linear layer & \texttt{layer1\_out\_features} & 320 \\
     Dimension of second linear layer & \texttt{layer2\_out\_features} & 640 \\
     Number of attention queries & \texttt{n\_attention\_queries} & 16 \\
     \bottomrule
\end{tabular}
\end{table}

\begin{table}[ht]
\caption{Aggregator model training parameters}
\label{table:training-params}
\centering
\begin{tabular}{l|r}
    \toprule
    Parameter name & Parameter value \\
    \midrule
    Batch size & 10 \\
    Optimizer & AdamW \cite{loshchilov2018decoupled} \\
    Learning rate & 0.0001 \\
    Weight decay & 0.05 \\
    Training loss & Binary cross-entropy \\
    \bottomrule
\end{tabular}
\end{table}

\subsubsection{Filtering criteria for biomarker performance categories}
\label{section:supplementary-methods-label-filters}

By assessing the \ac{AP} and \ac{AUROC} for \nvectorlabels{} biomarkers in the MSK test set, we observed a positive correlation between \ac{AP} and the prevalence of labels in the training set, defined as the proportion of positive samples (Figure~\ref{fig:S3}A). In contrast, \ac{AUROC} demonstrated no strong correlation with prevalence. To further evaluate model performance, we computed sensitivity and specificity for biomarkers with AUC~$>0.5$~(Figure~\ref{fig:S3}B).

Considering screening requirements, we categorize biomarkers into different groups by applying criteria defined in Table~\ref{table:filter-criteria}:
1. Baseline criteria for identifying biomarkers with signals exceeding those expected by chance
2. High-performing biomarkers as potential candidates for screening 
3. Top-performing biomarkers with high prevalence.
4. Hotspot detection

\begin{table}[htb]
    \caption{Filtering criteria}  
    \label{table:filter-criteria}
    \resizebox{\linewidth}{!}{
\centering
\begin{tabular}{l|l|l}
\toprule
Description & Parameter & $>$ threshold \\
\midrule
\multirow{3}{*}{Baseline (selected from all genes and all tumor types)} & AUC & 0.5 \\
 & positive ratio & 0.02 \\
 & N positives & 2 \\
\midrule
\multirow{6}{*}{High-performing biomarkers} & AUC & 0.75 \\
 & positive ratio & 0.02 \\
 & N positives & 2 \\
 & N total & 50 \\
 & sensitivity & 0.75 \\
 & specificity & 0.20 \\
\midrule
\multirow{4}{*}{Top-performing biomarkers with high prevalence} & AUC & 0.85 \\
 & positive ratio & 0.05 \\
 & sensitivity & 0.80 \\
 & specificity & 0.30 \\
 \midrule
\multirow{4}{*}{Hotspot detection} & AUC & 0.75 \\
 & N positives & 4 \\
 & N total & 50 \\
 & sensitivity & 0.75 \\
 & specificity & 0.20 \\
\bottomrule
\end{tabular}
}
\end{table}

\subsubsection{Single Label Aggregator Training}
\label{section:supplementary-methods-single-label}
The single label aggregator models were all trained with the model parameters described in Table~\ref{table:single-label-model-params}. The models were trained for 20 epochs with the same training parameters as OmniScreen, described in Table~\ref{table:training-params}. The input embedding (generated by Virchow2) consists of class token only (dimension 1280). The training time varied from 1h (KMT2D in esophagogastric cancer) to 8h (ESR1 in breast cancer) due to differing training set sizes. Training was performed on 1 NVIDIA A100 (40GB) node (8 GPUs).

\begin{table}[ht]
\caption{Single label aggregator model parameters}
\label{table:single-label-model-params}
\centering
\begin{tabular}{l|l|r}
     \toprule
     Parameter description & Parameter name & Parameter value \\
     \midrule
     Dimension of the input embedding & \texttt{in\_features} & 1280 \\
     Dimension of first linear layer & \texttt{layer1\_out\_features} & 160 \\
     Dimension of second linear layer & \texttt{layer2\_out\_features} & 320 \\
     Number of attention queries & \texttt{n\_attention\_queries} & 1 \\
     \bottomrule
\end{tabular}
\end{table}


\subsubsection{Biomarkers associated with targeted therapeutic hotspots}
\label{section:supplementary-biomarker-therapy-targets}
We evaluated OmniScreen's predictions of genomic alterations linked to therapeutic targets of \ac{FDA}-approved drugs across various cancer types. A list of 54 treatment-associated genes with specific hotspot mutations reported in \ac{MCG}~\cite{swanton2012my,holt2021my} and OncoKB~\cite{chakravarty2017oncokb,suehnholz2023cancer} was compiled, focusing on the actionable targets that are \ac{FDA}-approved predictive biomarkers (OncoKB therapeutic evidence level 1) or standard of care biomarkers recommended by the \ac{NCCN} or other expert panels (OncoKB therapeutic evidence level 2). The ground truth for each sample in the test set was determined based on the presence or absence of hotspot mutations specific to its cancer type.

We identified 52 predictive biomarkers among the 15 most common cancers where OmniScreen could predict the likelihood of the mutation, positioning it as a strong candidate for a screening assay (Figure~\ref{fig:S4}A, Table~\ref{table:results-hotspots}). For example, BRAF V600E mutation testing is standard of care in melanoma, glioma, thyroid, lung and \acp{CRC}, where such mutations are common and therapeutically actionable targets in these malignancies~\cite{di2023implications}. 

OmniScreen demonstrated strong performance in predicting BRAF mutations, with AUCs of 0.93 and 0.89 in primary and metastatic thyroid cancers, respectively. Additionally, it achieved notable accuracy in primary \acp{CRC} (AUC = 0.91), glioma (AUC = 0.88), and melanoma (AUC = 0.76), despite limited metastatic cases. 
In \ac{NSCLC}, where BRAF mutations are relatively rare (1-2\% of \ac{NSCLC}~\cite{planchard2024brafv600e}), our model accurately predicted the presence of the BRAF V600E mutation in confirmed cases, achieving an AUC of 0.77 (N=11/903, 1.2\%). Further improvements in predictive performance may be feasible with access to additional positive cases for model training.

OmniScreen achieved an AUC of 0.81 for predicting ERBB2 amplification (predictive of response to drugs such as trastuzumab) in both primary breast cancers and metastatic gastric cancers. 
In metastatic \ac{NSCLC}, our evaluation showed that the prediction of ERBB2 oncogenic alterations and amplification associated with response to Fam-Trastuzumab Deruxtecan-nxki both achieved an AUC of 0.79. In particular, the ERBB2 amplification prediction demonstrated a \ac{NPV} of 100\%, suggesting its potential to accurately identify patients without ERBB2 amplification who are unlikely to benefit from HER2-targeted therapy. 

In addition to ERBB2, OmniScreen identified multiple actionable oncogenic drivers in NSCLC (AUC $>$0.75) with approved \ac{TKI}-based therapies targeting specific mutations in genes such as EGFR, KRAS, MET, and RET.
In primary \ac{NSCLC}, OmniScreen demonstrated robust performance in predicting TKI-targetable EGFR mutations (p.L858R, p.T790M, exon 19 deletions, and exon 20 insertions; AUC=0.87), KRAS p.G12C (AUC=0.83), and MET exon 14 deletion/splicing mutations (AUC=0.79), highlighting its potential for precision oncology applications.
OmniScreen also performed well in detecting rarer \ac{NSCLC} mutations associated with targeted therapies, such as RET fusion detection, achieving an AUC of 0.77 in primary samples (N=12/903, 1.3\%) and 0.89 in metastatic samples (N=7/390, 1.8\%). Notably, RET fusion detection demonstrated an \ac{NPV} of 100\%, reliably excluding patients unlikely to harbor this alteration, and further underscoring the clinical utility of OmniScreen in screening both common and rare actionable mutations.

Other noteworthy examples include OmniScreen's prediction of FGFR3 hotspot mutations (R248C, S249C, G370C, Y373C), which are predictive of response to drugs like erdafitinib, reaching an AUC of 0.93 in primary bladder cancer, slightly better than its performance in metastatic lesions (AUC=0.82).

\subsubsection{Biomarkers associated with genome instability}
\label{section:supplementary-biomarker-genome-instability}
Genomic instability is a hallmark of many cancers and is associated with an increased likelihood of response to many therapies that improve immune system response to tumor cells. To evaluate whether phenotypes indicative of genomic instability could be detected, we examined three key measures: high tumor mutation burden (TMB-H), microsatellite instability or deficient mismatch repair (MSI-H/dMMR), and \acf{CIN} (Figure~\ref{fig:S4}C). 

OmniScreen demonstrated strong performance in predicting \ac{TMB-H}, achieving a mean AUC of 0.88 (range: 0.79-0.91) across nine cancer types. Notably, primary \acp{CRC} and endometrial cancers, both enriched in \ac{TMB-H} with prevalence~$>20$\%, demonstrated an AUC of 0.9. 
Similarly, the model performed well in metastatic esophagogastric cancer, achieving an AUC of 0.87.
In soft tissue sarcoma, where \ac{TMB-H} represents a rare molecular subgroup and the majority of cases exhibit low TMB levels, OmniScreen identified a small subset of primary samples with \ac{TMB-H} (N=8/346, 2.3\%) in our test set, achieving an AUC of 0.89. 
Further analysis of the TCGA evaluation set substantiated the efficacy of the model in detecting \ac{TMB-H} across five distinct cancer types, with a mean AUC of 0.89 (Table~\ref{table:results-merged}).

Our model exhibited exceptional diagnostic performance in identifying MSI-H/dMMR across various primary tumor types, including \ac{CRC} (AUC = 0.97), endometrial cancer (AUC = 0.87), esophagogastric cancer (AUC = 0.96), and bladder cancer (AUC = 0.97).
In colon adenocarcinomas, the highly attended tiles, to which OmniScreen assigned significant importance during inference, prominently displayed histopathological features characteristic of MSI-H/dMMR, including poorly differentiated, medullary-like patterns of tumor growth with ``pushing'' borders, tumor infiltrating lymphocytes (TILs), and mucin production (Figure~\ref{fig:S5}F). 
Analysis of the TCGA dataset further validated the robustness of OmniScreen in detecting MSI-H/dMMR phenotypes across \ac{CRC}, endometrial and gastric cancers, demonstrating a mean AUC of~0.9. 

Individuals with \ac{LS} are at an increased hereditary risk for MSI-H/dMMR-associated cancers~\cite{elze2023microsatellite}. \ac{LS} is diagnosed through the identification of germline pathogenic mutations in \ac{MMR} genes or EPCAM. Using a surrogate ground truth based on these mutations, OmniScreen achieved strong performance in primary \ac{CRC}, with an AUC of 0.85 for \ac{LS} prediction.

While MSI-H/dMMR highlights genomic instability driven by small-scale mutations, \ac{CIN} represents a distinct form of genomic instability characterized by large-scale structural and numerical chromosomal alterations. In this study, CIN data were exclusively curated for breast and ovarian cancers, and assessed using five metrics indicative of chromosomal aberrations: tetraploidy, \ac{WGD}, \ac{FGA}~$\geq30$\%, \ac{LOH}~$\geq50\%$, and \ac{GI index}~$\geq20\%$ (see Methods~\ref{section:methods}). Our model demonstrated robust performance in predicting \ac{CIN}, achieving a mean AUC of 0.84 in primary breast and ovarian samples. In metastatic lesions, slightly higher performance was observed in breast (AUC = 0.83) compared to ovarian cancer (AUC = 0.8).

\subsubsection{Biomarkers associated with histological subtypes}
\label{section:supplementary-biomarker-phenotype-association}
In renal cell carcinoma, loss of function in VHL is a hallmark of \ac{ccRCC}~\cite{hu2023tumor,zhang2024vhl}. The prediction of VHL oncogenic mutation accurately identified \ac{ccRCC} with an AUC of 0.94. The regions highly attended by OmniScreen demonstrated its ability to detect the characteristic clear cell histology of renal cell carcinoma, even in cases with high WHO/ISUP 4 nuclear grades (Figure~\ref{fig:S5}C).

Similarly, within pancreatic cancers, MEN1 and ATRX were frequently mutated in pancreatic neuroendocrine tumor~\cite{park2017daxx}. Prediction of oncogenic mutation/deletion in these two genes could diagnose pancreatic neuroendocrine tumor (MEN1, AUC=0.98; ATRX, AUC=0.93), while KRAS oncogenic mutation prediction correctly identified pancreatic adenocarcinoma diagnosis with an AUC of 0.95. Other examples of distinct phenotypic associations include the prediction of genomic alterations in GNAQ (AUC=0.92) and GNA11 (AUC=0.9), identifying uveal melanoma, and RET mutations detecting \ac{MTC} with an AUC of 0.99. 

OmniScreen effectively detected a wide range of genomic alterations and histopathological features across diverse soft tissue sarcoma subtypes, highlighting its versatility in diagnosing both common and rare sarcoma entities. For example, the prediction of RB1 deletion achieved high accuracy (AUC=0.91) in identifying leiomyosarcoma, a subtype of sarcoma with high incidence of RB1 loss. In RB1-loss WSIs, the model focused on hallmark features of leiomyosarcoma, such as spindle-shaped tumor cells with eosinophilic cytoplasm arranged in storiform or fascicular growth patterns and areas with pronounced nuclear atypia (Figure~\ref{fig:S5}D). OmniScreen accurately predicted MDM2 amplification to diagnose well-differentiated (WDLPS) and dedifferentiated liposarcoma (DDLPS) with AUCs of 0.91 and 0.86, respectively, reflecting the high prevalence of MDM2 amplification in these subtypes~\cite{gambella2023fish,sciot2021mdm2}. 
OmniScreen also effectively detected TERT oncogenic mutations for diagnosing myxoid liposarcoma (AUC=0.93), and demonstrated exceptional accuracy in predicting WT1 fusions to identify desmoplastic small round cell tumor (DSRCTs; AUC=0.99), which are canonically characterized by the EWSR1-WT1 fusion~\cite{wu2022multi}.

OmniScreen not only identifies established histological subtypes but also differentiates subsets within the same subtype based on distinct molecular profiles. In \ac{LUSC}, KMT2D deficiency promotes tumor progression~\cite{pan2023kmt2d} and enhances sensitivity to therapies targeting the RTK-RAS signaling pathway. Our evaluation demonstrated that diagnosing \ac{LUSC} through the prediction of KMT2D oncogenic mutation achieved an AUC of 0.90. Highly attended regions in KMT2D \ac{TP} WSIs indicated that the model focused on areas with moderate to high nuclear pleomorphism and atypia along with features characteristic of squamous cell differentiation, such as intercellular bridges and focal keratinization (Figure~\ref{fig:S5}E). KMT2D prediction by OmniScreen was also accurate in poorly differentiated examples, where squamous differentiation might require ancillary IHC to diagnose.
In contrast, SOX2 amplification represents a distinct category of chromosomal aberrations in \ac{LUSC}, contributing to tumor initiation and progression in a substantial subset of cases~\cite{liu2021predominant}. The prediction of SOX2 amplification achieved an AUC of 0.91 for diagnosing specific subsets of cases in \ac{LUSC}.

Lastly, a single biomarker can identify shared pathological features across tumor subtypes. For example, the prediction of ARID1A deleterious mutations, frequently found in clear cell and endometrioid ovarian cancers~\cite{takahashi2021treatment}, achieved AUCs of 0.95 and 0.89 for identifying \ac{OCCC} (N=29) and endometrioid ovarian cancers (N=23) respectively.

\subsubsection{Cost saving analysis}
\label{supplementary-method:costanalysis}

We first assumed a typical Phase 3 study seeking to enroll 500 patients with a mutation in a target gene in a particular type of cancer. We selected BRAF (\ac{CRC} and melanoma), EGFR (NSCLC), FGFR3 (bladder cancer), KRAS (NSCLC and \ac{CRC}), and MET (NSCLC), as these are commonly assessed by NGS or PCR clinically, allowing good estimates of the cost of these molecular assays along with a nominal cost of AI model screening. Prevalence of these mutations in these forms of cancer at published rates was assumed, with KRAS in NSCLC given a low estimate for a population with little to no smoking and a higher estimate for a population with high smoking prevalence. From this, we calculated the estimated cost savings due to reduced numbers of definitive NGS or PCR tests needed after our AI model identified cases unlikely to harbor the targeted mutation, and hence would not benefit from the targeted therapy or respond to targeted drugs.  

We found potential for substantial savings across all genes and cancer types investigated (Figure~\ref{fig:S4}B). Study enrollment savings for screening PCR or NGS were highest for genes in which the targeted mutation type was lower in prevalence, due to reduced numbers of mutation negative patients sent on for definitive molecular screening. On average, an 18\% and 26\% cost saving in our hypothetical setting could be achieved in PCR (range 5\%-30\%) and NGS (range 13\%-39\%) testing, respectively. This translates to an average cost reduction of approximately \$320K (range \$31K-\$1,215K) and \$1,086K (range \$188K-\$3,930K) for enrollment screening PCR and NGS testing.

The cost savings calculation for using an AI model to triage for downstream molecular testing 
were done according to the following method:

Assuming enrollment of a target number of patients all with tumors harboring a specific mutation, cost estimation for patient screening was calculated by 1) estimating cost to enroll the target number of patients using a conventional molecular test (NGS or PCR) only, and 2) NGS or PCR following AI-based pre-screening to eliminate patients who were not likely to have the targeted mutation. The cost savings was then calculated as follows:

\begin{itemize}
    \item Number of targeted enrolled patients: $N_{target}$
    \item Prevalence of a genomic alteration in a cancer: $prevalence$ 
    \item Sensitivity, i.e., the AI model algorithm's true positive rate: $sensitivity$
    \item Specificity, i.e., the AI model algorithm's true negative rate: $specificity$
    \item Cost of AI screening per patient: $C_{AI}$ 
    \item Cost of molecular testing per patient: $C_{testing}$
\end{itemize}

\begin{enumerate}

\item Cost estimation for patient screening using a conventional molecular testing:

Number of patients to be screened by molecular testing without AI model: 
\[N_{conventional} = \frac{N_{target}}{prevalence}\]

Total cost of patients to be screened by molecular testing only without engaging an AI model:
\[C_{conventional} = C_{testing} \times N_{conventional}\]

\item Cost estimation for patient screening using a molecular testing with an AI model for pre-screening:

Number of patients to be screened with an AI model: 
\[N_{screened} = \frac{N_{target}}{prevalence \times sensitivity}\]

Number of true positives (TP):
\[TP = N_{screened} \times prevalence \times sensitivity\]

Number of false positives (FP):
\[FP = N_{screened} \times (1 - prevalence) \times (1 - specificity)\]

Number of patients sent for molecular testing:
\[N_{sent} = TP + FP\]

Cost of AI model for screening all patients:
\[C_{total(AI)} = C_{AI} \times N_{screened} \]

Cost of molecular testing for patients sent for testing:
\[C_{total(testing)} = C_{testing} \times N_{sent} \]

Total cost of patients' molecular testing with AI screening: 
\[C_{with\_AI} = C_{total(AI)} + C_{total(testing)}\]

\end{enumerate}

Total cost saving by using an AI model for pre-screening:
\[C_{saving} = C_{conventional} - C_{with\_AI} \]

Hence, the percentage of cost reduction by using an AI model for pre-screening:
\[C\%= \frac{C_{saving}}{C_{conventional}} \]

\clearpage
\subsection{Supplementary Tables}
\label{section:supplementary-tables}

\begin{table}[ht]
\caption{{Genetic alteration biomarkers validated in both MSK and TCGA cohorts}}\label{table:results-merged}
\centering\resizebox{\linewidth}{!}{\begin{tabular}{lllrrrrrr}
\toprule
Gene & Biomarker & Cancer Type & \multicolumn{3}{r}{MSK} & \multicolumn{3}{r}{TCGA} \\
 &  &  & AUC & Sensitivity & Specificity & AUC & Sensitivity & Specificity \\
\midrule
E2F3 & E2F3 amplification & Bladder Cancer & 0.88 & 1.00 & 0.51 & 0.86 & 1.00 & 0.23 \\
FGFR & FGFR oncogenic mutation & Bladder Cancer & 0.92 & 0.95 & 0.47 & 0.80 & 0.86 & 0.46 \\
FGFR3 & FGFR3 oncogenic alterations: amplification/mutation & Bladder Cancer & 0.91 & 0.98 & 0.30 & 0.81 & 0.91 & 0.37 \\
 & ER & Breast Cancer & 0.95 & 0.93 & 0.76 & 0.92 & 0.93 & 0.79 \\
CDH1 & CDH1 oncogenic mutation & Breast Cancer & 0.94 & 0.87 & 0.91 & 0.88 & 0.72 & 0.88 \\
 & PR & Breast Cancer & 0.89 & 0.95 & 0.56 & 0.83 & 0.97 & 0.53 \\
\bottomrule
\end{tabular}

}
\end{table}

\begin{landscape}
\begingroup
\renewcommand{\arraystretch}{1.05}
\begin{table}[!ht]
\caption{Genetic alteration biomarkers associated with histologic subtypes}
\label{table:results-histology}
\centering
\resizebox{\linewidth}{!}{
\begin{tabular}{llllrrrrrrrrr}
\toprule
Cancer Type & Histology & Biomarker & KS-test adj.P-value & AUC & AP & Sensitivity & Specificity & PPV & NPV & Positive & Negative & Total \\
\midrule
Breast Cancer & Breast Invasive Lobular Carcinoma & CDH1 oncogenic alterations: deletion/mutation & 1.88e-74 & 0.94 & 0.68 & 0.86 & 0.90 & 0.51 & 0.98 & 131 & 1046 & 1177 \\
Colorectal Cancer & Mucinous Adenocarcinoma of the Colon and Rectum & GNAS oncogenic mutation & 6.98e-13 & 0.94 & 0.25 & 1.00 & 0.72 & 0.07 & 1.00 & 20 & 970 & 990 \\
Endometrial Cancer & Uterine Endometrioid Carcinoma & SOX17 oncogenic mutation & 2.15e-43 & 0.88 & 0.87 & 0.69 & 0.88 & 0.87 & 0.71 & 263 & 222 & 485 \\
Endometrial Cancer & Uterine Endometrioid Carcinoma & PTEN oncogenic alterations: deletion/mutation & 5.20e-46 & 0.88 & 0.88 & 0.86 & 0.74 & 0.80 & 0.82 & 263 & 222 & 485 \\
Endometrial Cancer & Uterine Endometrioid Carcinoma & CTCF oncogenic mutation & 7.92e-47 & 0.87 & 0.87 & 0.72 & 0.84 & 0.84 & 0.72 & 263 & 222 & 485 \\
Endometrial Cancer & Uterine Serous Carcinoma/Uterine Papillary Serous Carcinoma & TP53 oncogenic alterations: deletion/mutation & 1.04e-33 & 0.90 & 0.55 & 1.00 & 0.64 & 0.36 & 1.00 & 83 & 402 & 485 \\
Endometrial Cancer & Uterine Serous Carcinoma/Uterine Papillary Serous Carcinoma & CCNE1 oncogenic alterations: amplification/mutation & 8.86e-27 & 0.88 & 0.57 & 0.81 & 0.80 & 0.45 & 0.95 & 83 & 402 & 485 \\
\bottomrule
\end{tabular}
}
\end{table}
\endgroup
\end{landscape}

\begin{table}[!ht]
\caption{Genetic alteration biomarkers associated with specific therapeutic hotspot mutations in target genes.}
\label{table:results-hotspots}
\centering
\resizebox{\linewidth}{!}{
\begin{tabular}{llllrrrrrrrrr}
\toprule
Target Gene & Cancer Type & Tumor Site & Biomarker & AUC & AP & Sensitivity & Specificity & PPV & NPV & Positive & Negative & Total \\
\midrule
AKT1 & Bladder Cancer & Primary & AKT1 oncogenic mutation & 0.93 & 0.30 & 1.00 & 0.26 & 0.03 & 1.00 & 7 & 342 & 349 \\
AKT1 & Colorectal Cancer & Primary & AKT1 oncogenic mutation & 0.77 & 0.04 & 0.88 & 0.65 & 0.03 & 1.00 & 8 & 691 & 699 \\
AKT1 & Endometrial Cancer & Primary & AKT1 oncogenic mutation & 0.81 & 0.17 & 0.85 & 0.55 & 0.07 & 0.99 & 13 & 352 & 365 \\
AKT1 & Prostate Cancer & Metastasis & AKT1 oncogenic mutation & 0.77 & 0.07 & 1.00 & 0.24 & 0.03 & 1.00 & 6 & 291 & 297 \\
ATM & Colorectal Cancer & Primary & ATM oncogenic mutation & 0.77 & 0.24 & 0.90 & 0.34 & 0.09 & 0.98 & 48 & 651 & 699 \\
ATM & Melanoma & Metastasis & ATM deletion & 0.76 & 0.07 & 0.83 & 0.55 & 0.05 & 0.99 & 6 & 202 & 208 \\
ATR & Colorectal Cancer & Primary & ATR oncogenic mutation & 0.88 & 0.10 & 1.00 & 0.70 & 0.07 & 1.00 & 16 & 683 & 699 \\
\end{tabular}
}
\end{table}

\begin{table}\ContinuedFloat
\caption{Genetic alteration biomarkers associated with specific therapeutic hotspot mutations in target genes. (continued)}
\centering
\resizebox{\linewidth}{!}{
\begin{tabular}{llllrrrrrrrrr}
MLH1 & Colorectal Cancer & Primary & MLH1 oncogenic mutation & 0.95 & 0.42 & 1.00 & 0.78 & 0.10 & 1.00 & 17 & 682 & 699 \\
MLH1 & Endometrial Cancer & Primary & MLH1 oncogenic mutation & 0.86 & 0.24 & 0.82 & 0.74 & 0.09 & 0.99 & 11 & 354 & 365 \\
MSH2 & Bladder Cancer & Primary & MSH2 oncogenic mutation & 0.96 & 0.40 & 1.00 & 0.55 & 0.04 & 1.00 & 7 & 342 & 349 \\
MSH2 & Colorectal Cancer & Primary & MSH2 oncogenic mutation & 0.93 & 0.22 & 1.00 & 0.67 & 0.09 & 1.00 & 21 & 678 & 699 \\
MSH2 & Endometrial Cancer & Primary & MSH2 oncogenic mutation & 0.83 & 0.37 & 0.80 & 0.80 & 0.14 & 0.99 & 15 & 350 & 365 \\
MSH6 & Colorectal Cancer & Primary & MSH6 oncogenic mutation & 0.91 & 0.22 & 0.97 & 0.81 & 0.19 & 1.00 & 31 & 668 & 699 \\
NBN & Colorectal Cancer & Primary & NBN oncogenic mutation & 0.92 & 0.18 & 0.92 & 0.72 & 0.06 & 1.00 & 13 & 686 & 699 \\
NRAS & Soft Tissue Sarcoma & Primary & NRAS oncogenic mutation & 0.75 & 0.03 & 1.00 & 0.49 & 0.03 & 1.00 & 5 & 351 & 356 \\
\bottomrule
\end{tabular}
}
\end{table}

\begin{table}[hb]
\caption{Genes candidates of 5 canonical signaling pathways (DDR, RTK, HRD, mTOR, TGF-$\beta$).}
\label{table:msk-pathways}
\resizebox{\linewidth}{!}{
\centering
\renewcommand{\arraystretch}{1.5}
\begin{tabular}{l p{9cm}}
\toprule
Pathway & Genes \\
\midrule
DNA damage response (DDR) (N=23) & ATM, ATR, ATRX, BRCA1, BRCA2, BRIP1, FANCA, FANCC, FANCD2, FANCE, FANCF, FANCG, FANCL, MDM2, MDM4, MLH1, MUTYH, NPM1, PALB2, PPP2R1A, RAD50, RAD51, STAG2 \\
Receptor tyrosine kinase (RTK) (N=34) & ALK, CBL, CSF1R, DDR2, EGFR, EPHA3, EPHA5, EPHB1, ERBB2, ERBB3, ERBB4, FGF19, FGF3, FGF4, FGFR1, FGFR2, FGFR3, FGFR4, FLT1, FLT3, FLT4, HGF, IGF1R, KIT, MET, NF1, NTRK1, NTRK2, NTRK3, PDGFRA, PDGFRB, PTPN11, RET, ROS1 \\
Homologous recombination deficiency (HRD) (N=11) & BRCA1, BRCA2, PALB2, ATM, CHEK1, CHEK2, RAD51, FANCA, CDK12, RAD51B, RAD51C \\
mTOR signaling (N=16) & AKT1, AKT2, AKT3, CRKL, IRS2, MTOR, PIK3CA, PIK3CG, PIK3R1, PIK3R2, PTEN, RICTOR, RNF43, RPTOR, TSC1, TSC2 \\
TGF-$\beta$ signaling (N=5) & SMAD2, SMAD3, SMAD4, TGFBR1, TGFBR2 \\
\bottomrule
\end{tabular}
\renewcommand{\arraystretch}{1}
}
\end{table}

\begin{table}[!ht]
\caption{Performance Metrics for Biomarker Prediction on the MSK Test Set. The performance of biomarkers in primary and metastatic lesions of the 15 most common cancer types are summarized in the following sub-tables \ref{table:results-bladder-cancer} - \ref{table:results-soft-tissue-sarcoma}. Metrics include Area Under the Curve (AUC), Average Precision (AP), Sensitivity, Specificity, Positive Predictive Value (PPV), and Negative Predictive Value (NPV). Sample counts are provided for each biomarker, indicating the number of positive, negative, and total samples.}
\label{table:msk-evaluation}
\begin{subtable}[h]{\linewidth}
\caption{Biomarkers detected in Bladder Cancer}\label{table:results-bladder-cancer}\centering\resizebox{\linewidth}{!}{\begin{tabular}{llrrrrrrrrr}
\toprule
Tumor Site & Biomarker & AUC & AP & Sensitivity & Specificity & PPV & NPV & Positive & Negative & Total \\
\midrule
Primary & MSI-H vs MSS & 0.97 & 0.60 & 1.00 & 0.61 & 0.06 & 1.00 & 8 & 305 & 313 \\
Primary & MSH2 oncogenic alterations: deletion/mutation & 0.97 & 0.40 & 1.00 & 0.64 & 0.05 & 1.00 & 7 & 342 & 349 \\
Primary & MSI-H/dMMR & 0.97 & 0.46 & 1.00 & 0.60 & 0.06 & 1.00 & 8 & 305 & 313 \\
Primary & MSH2 oncogenic mutation & 0.96 & 0.40 & 1.00 & 0.55 & 0.04 & 1.00 & 7 & 342 & 349 \\
Primary & FGFR3 hotspots/FGFR fusions & 0.94 & 0.86 & 1.00 & 0.35 & 0.32 & 1.00 & 76 & 251 & 327 \\
Primary & AKT1 oncogenic mutation & 0.93 & 0.30 & 1.00 & 0.26 & 0.03 & 1.00 & 7 & 342 & 349 \\
Primary & FGFR3 hotspots & 0.93 & 0.86 & 0.97 & 0.23 & 0.28 & 0.97 & 76 & 252 & 328 \\
\end{tabular}

}
\end{subtable}
\end{table}
 
\begin{table}\ContinuedFloat
\addtocounter{subtable}{-1}
\begin{subtable}[h]{\linewidth}
\caption{Biomarkers detected in Bladder Cancer (continued)}
\centering\resizebox{\linewidth}{!}{\begin{tabular}{llrrrrrrrrr}
Tumor Site & Biomarker & AUC & AP & Sensitivity & Specificity & PPV & NPV & Positive & Negative & Total \\
\midrule
Metastasis & CYSLTR2 deletion & 1.00 & 1.00 & 1.00 & 0.49 & 0.08 & 1.00 & 3 & 67 & 70 \\
Metastasis & RB1 deletion & 1.00 & 1.00 & 1.00 & 0.80 & 0.15 & 1.00 & 3 & 87 & 90 \\
Metastasis & TGFBR1 oncogenic alterations: deletion/mutation & 0.99 & 0.83 & 1.00 & 0.72 & 0.07 & 1.00 & 2 & 88 & 90 \\
Metastasis & NFE2L2 oncogenic mutation & 0.98 & 0.70 & 1.00 & 0.44 & 0.06 & 1.00 & 3 & 87 & 90 \\
Metastasis & FOXO1 deletion & 0.97 & 0.33 & 1.00 & 0.78 & 0.10 & 1.00 & 2 & 88 & 90 \\
Metastasis & SDHA amplification & 0.96 & 0.60 & 1.00 & 0.49 & 0.08 & 1.00 & 4 & 86 & 90 \\
Metastasis & TGFBR1 deletion & 0.94 & 0.27 & 1.00 & 0.90 & 0.18 & 1.00 & 2 & 88 & 90 \\
\bottomrule
\end{tabular}

}
\end{subtable}
\end{table} 

\begin{table}\ContinuedFloat
\begin{subtable}[h]{\linewidth}
\caption{Biomarkers detected in Breast Cancer}\label{table:results-breast-cancer}\centering\resizebox{\linewidth}{!}{\begin{tabular}{llrrrrrrrrr}
\toprule
Tumor Site & Biomarker & AUC & AP & Sensitivity & Specificity & PPV & NPV & Positive & Negative & Total \\
\midrule
Primary & ER & 0.95 & 0.99 & 0.93 & 0.76 & 0.94 & 0.75 & 326 & 87 & 413 \\
Primary & ER-/PR-/HER2- & 0.94 & 0.68 & 0.94 & 0.81 & 0.51 & 0.99 & 68 & 325 & 393 \\
Primary & PREX2 amplification & 0.94 & 0.23 & 1.00 & 0.78 & 0.10 & 1.00 & 11 & 437 & 448 \\
Primary & CDH1 oncogenic mutation & 0.94 & 0.65 & 0.87 & 0.91 & 0.52 & 0.98 & 54 & 481 & 535 \\
Primary & ER-/PR-/HER2+ & 0.94 & 0.34 & 0.82 & 0.89 & 0.18 & 0.99 & 11 & 382 & 393 \\
Primary & CDH1 oncogenic alterations: deletion/mutation & 0.94 & 0.64 & 0.89 & 0.91 & 0.53 & 0.99 & 56 & 479 & 535 \\
Primary & RB1 oncogenic alterations: deletion/mutation & 0.92 & 0.21 & 1.00 & 0.44 & 0.05 & 1.00 & 16 & 519 & 535 \\
\end{tabular}

}
\end{subtable}
\end{table}

\begin{table}\ContinuedFloat
\addtocounter{subtable}{-1}
\begin{subtable}[h]{\linewidth}
\caption{Biomarkers detected in Breast Cancer (continued)}
\centering\resizebox{\linewidth}{!}{\begin{tabular}{llrrrrrrrrr}
Tumor Site & Biomarker & AUC & AP & Sensitivity & Specificity & PPV & NPV & Positive & Negative & Total \\
\midrule
Metastasis & ER-/PR-/HER2- & 0.87 & 0.60 & 0.95 & 0.59 & 0.36 & 0.98 & 78 & 327 & 405 \\
Metastasis & ER & 0.86 & 0.95 & 0.86 & 0.69 & 0.90 & 0.61 & 340 & 110 & 450 \\
Metastasis & Tetraploidy & 0.86 & 0.80 & 0.96 & 0.48 & 0.56 & 0.94 & 166 & 240 & 406 \\
Metastasis & Whole genome doubling & 0.84 & 0.60 & 0.89 & 0.63 & 0.39 & 0.95 & 61 & 234 & 295 \\
Metastasis & DUSP4 deletion & 0.84 & 0.36 & 0.80 & 0.75 & 0.09 & 0.99 & 15 & 502 & 517 \\
Metastasis & HER2 Amplification & 0.83 & 0.47 & 0.94 & 0.29 & 0.15 & 0.97 & 53 & 394 & 447 \\
Metastasis & RAD52 amplification & 0.83 & 0.11 & 1.00 & 0.46 & 0.04 & 1.00 & 13 & 629 & 642 \\
\bottomrule
\end{tabular}

}
\end{subtable}
\end{table}

\begin{table}\ContinuedFloat
\begin{subtable}[h]{\linewidth}
\caption{Biomarkers detected in Colorectal Cancer}\label{table:results-colorectal-cancer}\centering\resizebox{\linewidth}{!}{\begin{tabular}{llrrrrrrrrr}
\toprule
Tumor Site & Biomarker & AUC & AP & Sensitivity & Specificity & PPV & NPV & Positive & Negative & Total \\
\midrule
Primary & MSI-H vs MSS & 0.98 & 0.91 & 0.93 & 0.93 & 0.72 & 0.99 & 105 & 563 & 668 \\
Primary & MSI-H/dMMR & 0.97 & 0.91 & 0.93 & 0.92 & 0.69 & 0.99 & 106 & 562 & 668 \\
Primary & HNF1A oncogenic alterations: deletion/mutation & 0.95 & 0.39 & 0.84 & 0.89 & 0.21 & 0.99 & 25 & 674 & 699 \\
Primary & TGFBR2 oncogenic mutation & 0.95 & 0.50 & 0.95 & 0.81 & 0.22 & 1.00 & 38 & 661 & 699 \\
Primary & MLH1 oncogenic mutation & 0.95 & 0.42 & 1.00 & 0.78 & 0.10 & 1.00 & 17 & 682 & 699 \\
Primary & SPEN oncogenic alterations: deletion/mutation & 0.95 & 0.20 & 0.93 & 0.86 & 0.12 & 1.00 & 14 & 685 & 699 \\
Primary & HNF1A oncogenic mutation & 0.95 & 0.40 & 0.80 & 0.89 & 0.21 & 0.99 & 25 & 674 & 699 \\
\end{tabular}

}
\end{subtable}
\end{table}

\begin{table}\ContinuedFloat
\addtocounter{subtable}{-1}
\begin{subtable}[h]{\linewidth}
\caption{Biomarkers detected in Colorectal Cancer (continued)}
\centering\resizebox{\linewidth}{!}{\begin{tabular}{llrrrrrrrrr}
Tumor Site & Biomarker & AUC & AP & Sensitivity & Specificity & PPV & NPV & Positive & Negative & Total \\
\midrule
Primary & JAK1 oncogenic mutation & 0.87 & 0.13 & 0.90 & 0.81 & 0.12 & 1.00 & 20 & 679 & 699 \\
Primary & INPPL1 oncogenic alterations: deletion/mutation & 0.87 & 0.12 & 0.93 & 0.74 & 0.08 & 1.00 & 15 & 612 & 627 \\
Primary & AURKA amplification & 0.86 & 0.24 & 0.82 & 0.81 & 0.10 & 0.99 & 17 & 682 & 699 \\
Primary & PTEN oncogenic mutation & 0.86 & 0.35 & 0.95 & 0.58 & 0.13 & 0.99 & 42 & 657 & 699 \\
Primary & RTEL1 amplification & 0.86 & 0.30 & 0.82 & 0.72 & 0.15 & 0.99 & 34 & 593 & 627 \\
Primary & GNAS amplification & 0.86 & 0.23 & 0.84 & 0.72 & 0.10 & 0.99 & 25 & 674 & 699 \\
Primary & AGO2 amplification & 0.86 & 0.30 & 0.94 & 0.52 & 0.05 & 1.00 & 16 & 611 & 627 \\
\end{tabular}

}
\end{subtable}
\end{table}

\begin{table}\ContinuedFloat
\addtocounter{subtable}{-1}
\begin{subtable}[h]{\linewidth}
\caption{Biomarkers detected in Colorectal Cancer (continued)}
\centering\resizebox{\linewidth}{!}{\begin{tabular}{llrrrrrrrrr}
Tumor Site & Biomarker & AUC & AP & Sensitivity & Specificity & PPV & NPV & Positive & Negative & Total \\
\midrule
Metastasis & KMT2B oncogenic mutation & 0.95 & 0.48 & 1.00 & 0.74 & 0.10 & 1.00 & 6 & 211 & 217 \\
Metastasis & MSI-H/dMMR & 0.93 & 0.58 & 0.83 & 0.98 & 0.45 & 1.00 & 6 & 250 & 256 \\
Metastasis & KMT2B oncogenic alterations: deletion/mutation & 0.92 & 0.54 & 0.83 & 0.70 & 0.07 & 0.99 & 6 & 211 & 217 \\
Metastasis & MALT1 deletion & 0.92 & 0.33 & 1.00 & 0.32 & 0.06 & 1.00 & 11 & 267 & 278 \\
Metastasis & DIS3 amplification & 0.92 & 0.21 & 1.00 & 0.63 & 0.05 & 1.00 & 6 & 285 & 291 \\
Metastasis & TOP1 amplification & 0.92 & 0.28 & 0.89 & 0.77 & 0.11 & 1.00 & 9 & 282 & 291 \\
Metastasis & LATS2 amplification & 0.91 & 0.43 & 1.00 & 0.46 & 0.07 & 1.00 & 11 & 280 & 291 \\
\bottomrule
\end{tabular}

}
\end{subtable}
\end{table}
 
\begin{table}\ContinuedFloat
\begin{subtable}[h]{\linewidth}
\caption{Biomarkers detected in Endometrial Cancer}\label{table:results-endometrial-cancer}\centering\resizebox{\linewidth}{!}{\begin{tabular}{llrrrrrrrrr}
\toprule
Tumor Site & Biomarker & AUC & AP & Sensitivity & Specificity & PPV & NPV & Positive & Negative & Total \\
\midrule
Primary & JAK1 oncogenic alterations: amplification/mutation & 0.91 & 0.55 & 0.76 & 0.78 & 0.23 & 0.97 & 29 & 336 & 365 \\
Primary & JAK1 oncogenic alterations: deletion/mutation & 0.91 & 0.55 & 0.83 & 0.80 & 0.26 & 0.98 & 29 & 336 & 365 \\
Primary & JAK1 oncogenic mutation & 0.91 & 0.52 & 0.76 & 0.82 & 0.26 & 0.98 & 29 & 336 & 365 \\
Primary & MSI-H vs MSS & 0.91 & 0.59 & 0.85 & 0.82 & 0.40 & 0.97 & 39 & 283 & 322 \\
Primary & TP53 oncogenic alterations: deletion/mutation & 0.90 & 0.88 & 0.85 & 0.86 & 0.82 & 0.89 & 154 & 211 & 365 \\
Primary & TMB-H & 0.90 & 0.80 & 0.85 & 0.76 & 0.58 & 0.93 & 101 & 264 & 365 \\
Primary & TP53 oncogenic mutation & 0.90 & 0.87 & 0.87 & 0.87 & 0.83 & 0.90 & 154 & 211 & 365 \\
\end{tabular}

}
\end{subtable}
\end{table}

\begin{table}\ContinuedFloat
\addtocounter{subtable}{-1}
\begin{subtable}[h]{\linewidth}
\caption{Biomarkers detected in Endometrial Cancer (continued)}
\centering\resizebox{\linewidth}{!}{\begin{tabular}{llrrrrrrrrr}
Tumor Site & Biomarker & AUC & AP & Sensitivity & Specificity & PPV & NPV & Positive & Negative & Total \\
\midrule
Metastasis & MSI-H vs MSS & 0.96 & 0.73 & 1.00 & 0.90 & 0.57 & 1.00 & 12 & 94 & 106 \\
Metastasis & CTNNB1 oncogenic alterations: amplification/mutation & 0.95 & 0.86 & 0.97 & 0.82 & 0.64 & 0.99 & 29 & 91 & 120 \\
Metastasis & CTNNB1 oncogenic mutation & 0.95 & 0.85 & 0.97 & 0.81 & 0.62 & 0.99 & 29 & 91 & 120 \\
Metastasis & MSH6 oncogenic mutation & 0.95 & 0.48 & 1.00 & 0.88 & 0.18 & 1.00 & 3 & 117 & 120 \\
Metastasis & MSI-H/dMMR & 0.93 & 0.67 & 0.85 & 0.89 & 0.52 & 0.98 & 13 & 94 & 107 \\
Metastasis & RNF43 oncogenic mutation & 0.93 & 0.58 & 1.00 & 0.80 & 0.27 & 1.00 & 8 & 112 & 120 \\
Metastasis & TP53 oncogenic alterations: deletion/mutation & 0.93 & 0.94 & 0.88 & 0.77 & 0.81 & 0.84 & 64 & 56 & 120 \\
\bottomrule
\end{tabular}

}
\end{subtable}
\end{table}

\begin{table}\ContinuedFloat
\begin{subtable}[h]{\linewidth}
\caption{Biomarkers detected in Esophagogastric Cancer}\label{table:results-esophagogastric-cancer}\centering\resizebox{\linewidth}{!}{\begin{tabular}{llrrrrrrrrr}
\toprule
Tumor Site & Biomarker & AUC & AP & Sensitivity & Specificity & PPV & NPV & Positive & Negative & Total \\
\midrule
Primary & SOX2 amplification & 0.97 & 0.76 & 0.83 & 0.96 & 0.29 & 1.00 & 6 & 284 & 290 \\
Primary & MSH3 oncogenic mutation & 0.97 & 0.33 & 1.00 & 0.88 & 0.15 & 1.00 & 5 & 239 & 244 \\
Primary & MSH3 oncogenic alterations: deletion/mutation & 0.97 & 0.44 & 1.00 & 0.89 & 0.16 & 1.00 & 5 & 239 & 244 \\
Primary & MSI-H vs MSS & 0.96 & 0.84 & 0.92 & 0.90 & 0.31 & 1.00 & 12 & 255 & 267 \\
Primary & MSI-H/dMMR & 0.96 & 0.84 & 1.00 & 0.54 & 0.10 & 1.00 & 13 & 254 & 267 \\
Primary & VEGFA oncogenic alterations: amplification/mutation & 0.92 & 0.26 & 1.00 & 0.38 & 0.07 & 1.00 & 13 & 277 & 290 \\
Primary & VEGFA amplification & 0.91 & 0.26 & 1.00 & 0.38 & 0.07 & 1.00 & 13 & 277 & 290 \\
\end{tabular}

}
\end{subtable}
\end{table}

\begin{table}\ContinuedFloat
\addtocounter{subtable}{-1}
\begin{subtable}[h]{\linewidth}
\caption{Biomarkers detected in Esophagogastric Cancer (continued)}
\centering\resizebox{\linewidth}{!}{\begin{tabular}{llrrrrrrrrr}
Tumor Site & Biomarker & AUC & AP & Sensitivity & Specificity & PPV & NPV & Positive & Negative & Total \\
\midrule
Metastasis & STK11 oncogenic alterations: deletion/mutation & 0.99 & 0.83 & 1.00 & 0.60 & 0.06 & 1.00 & 2 & 82 & 84 \\
Metastasis & KMT2A amplification & 0.99 & 0.76 & 1.00 & 0.47 & 0.07 & 1.00 & 3 & 81 & 84 \\
Metastasis & MSI-H vs MSS & 0.98 & 0.50 & 1.00 & 0.93 & 0.29 & 1.00 & 2 & 69 & 71 \\
Metastasis & PAK5 amplification & 0.98 & 0.42 & 1.00 & 0.76 & 0.09 & 1.00 & 2 & 82 & 84 \\
Metastasis & PMAIP1 oncogenic alterations: deletion/mutation & 0.97 & 0.48 & 1.00 & 0.74 & 0.12 & 1.00 & 3 & 81 & 84 \\
Metastasis & MSI-H/dMMR & 0.97 & 0.45 & 1.00 & 0.64 & 0.07 & 1.00 & 2 & 69 & 71 \\
Metastasis & TERT oncogenic mutation & 0.96 & 0.43 & 1.00 & 0.38 & 0.06 & 1.00 & 3 & 81 & 84 \\
\bottomrule
\end{tabular}

}
\end{subtable}
\end{table}

\begin{table}\ContinuedFloat
\begin{subtable}[h]{\linewidth}
\caption{Biomarkers detected in Glioma}\label{table:results-glioma}\centering\resizebox{\linewidth}{!}{\begin{tabular}{llrrrrrrrrr}
\toprule
Tumor Site & Biomarker & AUC & AP & Sensitivity & Specificity & PPV & NPV & Positive & Negative & Total \\
\midrule
Primary & CIC oncogenic mutation & 0.96 & 0.59 & 0.82 & 0.92 & 0.38 & 0.99 & 22 & 362 & 384 \\
Primary & CIC oncogenic alterations: deletion/mutation & 0.96 & 0.53 & 0.82 & 0.94 & 0.45 & 0.99 & 22 & 362 & 384 \\
Primary & IDH1 oncogenic alterations: amplification/mutation & 0.95 & 0.82 & 0.89 & 0.86 & 0.71 & 0.96 & 104 & 280 & 384 \\
Primary & RET deletion & 0.95 & 0.41 & 0.95 & 0.85 & 0.25 & 1.00 & 19 & 365 & 384 \\
Primary & FGFR3 fusion & 0.95 & 0.42 & 1.00 & 0.49 & 0.05 & 1.00 & 11 & 373 & 384 \\
Primary & IDH1 oncogenic mutation & 0.94 & 0.81 & 0.88 & 0.87 & 0.72 & 0.95 & 104 & 280 & 384 \\
Primary & NOTCH1 oncogenic mutation & 0.93 & 0.32 & 1.00 & 0.78 & 0.12 & 1.00 & 11 & 373 & 384 \\
\bottomrule
\end{tabular}

}
\end{subtable}
\end{table}

\begin{table}\ContinuedFloat
\begin{subtable}[h]{\linewidth}
\caption{Biomarkers detected in Hepatobiliary Cancer}\label{table:results-hepatobiliary-cancer}\centering\resizebox{\linewidth}{!}{\begin{tabular}{llrrrrrrrrr}
\toprule
Tumor Site & Biomarker & AUC & AP & Sensitivity & Specificity & PPV & NPV & Positive & Negative & Total \\
\midrule
Primary & CCNE1 amplification & 0.98 & 0.46 & 1.00 & 0.64 & 0.06 & 1.00 & 4 & 165 & 169 \\
Primary & CCNE1 oncogenic alterations: amplification/mutation & 0.98 & 0.41 & 1.00 & 0.70 & 0.07 & 1.00 & 4 & 165 & 169 \\
Primary & STK11 oncogenic alterations: deletion/mutation & 0.95 & 0.28 & 1.00 & 0.86 & 0.15 & 1.00 & 4 & 165 & 169 \\
Primary & SMAD4 oncogenic mutation & 0.93 & 0.46 & 1.00 & 0.58 & 0.13 & 1.00 & 10 & 159 & 169 \\
Primary & SMAD4 oncogenic alterations: deletion/mutation & 0.93 & 0.44 & 1.00 & 0.59 & 0.15 & 1.00 & 11 & 158 & 169 \\
Primary & KRAS oncogenic mutation & 0.92 & 0.52 & 1.00 & 0.57 & 0.14 & 1.00 & 11 & 158 & 169 \\
Primary & IDH2 oncogenic mutation & 0.92 & 0.59 & 1.00 & 0.37 & 0.07 & 1.00 & 8 & 161 & 169 \\
\bottomrule
\end{tabular}

}
\end{subtable} 
\end{table}

\begin{table}\ContinuedFloat
\begin{subtable}[h]{\linewidth}
\caption{Biomarkers detected in Melanoma}\label{table:results-melanoma}\centering\resizebox{\linewidth}{!}{\begin{tabular}{llrrrrrrrrr}
\toprule
Tumor Site & Biomarker & AUC & AP & Sensitivity & Specificity & PPV & NPV & Positive & Negative & Total \\
\midrule
Primary & GNA11 oncogenic alterations: amplification/mutation & 0.97 & 0.56 & 1.00 & 0.73 & 0.09 & 1.00 & 3 & 118 & 121 \\
Primary & GNA11 oncogenic mutation & 0.96 & 0.52 & 1.00 & 0.81 & 0.12 & 1.00 & 3 & 118 & 121 \\
Primary & RECQL4 amplification & 0.95 & 0.49 & 1.00 & 0.54 & 0.07 & 1.00 & 4 & 117 & 121 \\
Primary & BAP1 oncogenic alterations: deletion/mutation & 0.94 & 0.33 & 1.00 & 0.56 & 0.05 & 1.00 & 3 & 118 & 121 \\
Primary & AGO2 amplification & 0.94 & 0.47 & 1.00 & 0.52 & 0.08 & 1.00 & 4 & 98 & 102 \\
Primary & PREX2 amplification & 0.94 & 0.61 & 1.00 & 0.48 & 0.06 & 1.00 & 3 & 99 & 102 \\
Primary & SF3B1 oncogenic mutation & 0.93 & 0.51 & 1.00 & 0.60 & 0.19 & 1.00 & 10 & 111 & 121 \\
\bottomrule
\end{tabular}

}
\end{subtable}
\end{table}

\begin{table}\ContinuedFloat
\begin{subtable}[h]{\linewidth}
\caption{Biomarkers detected in Non-Small Cell Lung Cancer}\label{table:results-non-small-cell-lung-cancer}\centering\resizebox{\linewidth}{!}{\begin{tabular}{llrrrrrrrrr}
\toprule
Tumor Site & Biomarker & AUC & AP & Sensitivity & Specificity & PPV & NPV & Positive & Negative & Total \\
\midrule
Primary & SOX2 amplification & 0.96 & 0.40 & 0.86 & 0.93 & 0.22 & 1.00 & 21 & 882 & 903 \\
Primary & PRKCI amplification & 0.92 & 0.23 & 0.88 & 0.81 & 0.09 & 1.00 & 16 & 742 & 758 \\
Primary & FGFR1 amplification & 0.90 & 0.10 & 1.00 & 0.64 & 0.06 & 1.00 & 19 & 884 & 903 \\
Primary & FGFR1 oncogenic alterations: amplification/mutation & 0.89 & 0.09 & 1.00 & 0.69 & 0.06 & 1.00 & 19 & 884 & 903 \\
Primary & RB1 oncogenic alterations: deletion/mutation & 0.89 & 0.23 & 1.00 & 0.51 & 0.07 & 1.00 & 34 & 869 & 903 \\
Primary & EGFR oncogenic mutation & 0.88 & 0.67 & 0.94 & 0.58 & 0.37 & 0.97 & 187 & 716 & 903 \\
Primary & RB1 oncogenic mutation & 0.87 & 0.22 & 0.97 & 0.56 & 0.08 & 1.00 & 33 & 870 & 903 \\
\bottomrule
\end{tabular}

}
\end{subtable}
\end{table}

\begin{table}\ContinuedFloat
\begin{subtable}[h]{\linewidth}
\caption{Biomarkers detected in Ovarian Cancer}\label{table:results-ovarian-cancer}\centering\resizebox{\linewidth}{!}{\begin{tabular}{llrrrrrrrrr}
\toprule
Tumor Site & Biomarker & AUC & AP & Sensitivity & Specificity & PPV & NPV & Positive & Negative & Total \\
\midrule
Primary & CTNNB1 oncogenic alterations: amplification/mutation & 0.99 & 0.82 & 1.00 & 0.75 & 0.13 & 1.00 & 8 & 209 & 217 \\
Primary & CTNNB1 oncogenic mutation & 0.99 & 0.79 & 1.00 & 0.78 & 0.15 & 1.00 & 8 & 209 & 217 \\
Primary & KMT2B oncogenic alterations: deletion/mutation & 0.97 & 0.36 & 1.00 & 0.93 & 0.24 & 1.00 & 4 & 192 & 196 \\
Primary & KMT2B oncogenic mutation & 0.97 & 0.37 & 0.75 & 0.96 & 0.27 & 0.99 & 4 & 192 & 196 \\
Primary & TP53 oncogenic mutation & 0.96 & 0.97 & 0.97 & 0.82 & 0.91 & 0.94 & 143 & 74 & 217 \\
Primary & TP53 oncogenic alterations: deletion/mutation & 0.96 & 0.97 & 0.96 & 0.82 & 0.91 & 0.91 & 143 & 74 & 217 \\
Primary & CTCF oncogenic alterations: deletion/mutation & 0.96 & 0.39 & 1.00 & 0.60 & 0.06 & 1.00 & 5 & 212 & 217 \\
\end{tabular}

}
\end{subtable}
\end{table}

\begin{table}\ContinuedFloat
\addtocounter{subtable}{-1}
\begin{subtable}[h]{\linewidth}
\caption{Biomarkers detected in Ovarian Cancer (continued)}
\centering\resizebox{\linewidth}{!}{\begin{tabular}{llrrrrrrrrr}
Tumor Site & Biomarker & AUC & AP & Sensitivity & Specificity & PPV & NPV & Positive & Negative & Total \\
\midrule
Metastasis & SPRTN amplification & 1.00 & 1.00 & 1.00 & 0.84 & 0.12 & 1.00 & 2 & 86 & 88 \\
Metastasis & RBM10 deletion & 0.98 & 0.41 & 1.00 & 0.74 & 0.08 & 1.00 & 5 & 227 & 232 \\
Metastasis & MTAP oncogenic alterations: deletion/mutation & 0.96 & 0.60 & 1.00 & 0.33 & 0.03 & 1.00 & 2 & 86 & 88 \\
Metastasis & TP53 oncogenic alterations: deletion/mutation & 0.95 & 0.99 & 0.94 & 0.77 & 0.95 & 0.73 & 193 & 39 & 232 \\
Metastasis & AR deletion & 0.95 & 0.23 & 1.00 & 0.79 & 0.09 & 1.00 & 5 & 227 & 232 \\
Metastasis & TP53 oncogenic mutation & 0.94 & 0.99 & 0.93 & 0.77 & 0.95 & 0.70 & 193 & 39 & 232 \\
Metastasis & PGBD5 amplification & 0.94 & 0.58 & 1.00 & 0.77 & 0.09 & 1.00 & 2 & 86 & 88 \\
\bottomrule
\end{tabular}

}
\end{subtable}
\end{table}

\begin{table}\ContinuedFloat
\begin{subtable}[h]{\linewidth}
\caption{Biomarkers detected in Pancreatic Cancer}\label{table:results-pancreatic-cancer}\centering\resizebox{\linewidth}{!}{\begin{tabular}{llrrrrrrrrr}
\toprule
Tumor Site & Biomarker & AUC & AP & Sensitivity & Specificity & PPV & NPV & Positive & Negative & Total \\
\midrule
Primary & MEN1 oncogenic mutation & 0.97 & 0.63 & 1.00 & 0.93 & 0.47 & 1.00 & 14 & 231 & 245 \\
Primary & MEN1 oncogenic alterations: deletion/mutation & 0.97 & 0.59 & 1.00 & 0.93 & 0.45 & 1.00 & 14 & 231 & 245 \\
Primary & MTAP oncogenic alterations: deletion/mutation & 0.93 & 0.37 & 1.00 & 0.78 & 0.20 & 1.00 & 3 & 55 & 58 \\
Primary & MTAP deletion & 0.92 & 0.34 & 1.00 & 0.85 & 0.27 & 1.00 & 3 & 55 & 58 \\
Primary & TMB-H & 0.91 & 0.60 & 0.83 & 0.63 & 0.05 & 0.99 & 6 & 233 & 239 \\
Primary & GNAS oncogenic alterations: amplification/mutation & 0.88 & 0.45 & 1.00 & 0.37 & 0.07 & 1.00 & 11 & 234 & 245 \\
Primary & KRAS oncogenic mutation & 0.88 & 0.95 & 0.91 & 0.69 & 0.91 & 0.67 & 191 & 54 & 245 \\
\bottomrule
\end{tabular}

}
\end{subtable}

\begin{subtable}[h]{\linewidth}
\caption{Biomarkers detected in Prostate Cancer}\label{table:results-prostate-cancer}\centering\resizebox{\linewidth}{!}{\begin{tabular}{llrrrrrrrrr}
\toprule
Tumor Site & Biomarker & AUC & AP & Sensitivity & Specificity & PPV & NPV & Positive & Negative & Total \\
\midrule
Primary & PREX2 amplification & 0.97 & 0.49 & 1.00 & 0.64 & 0.06 & 1.00 & 7 & 296 & 303 \\
Primary & PRDM14 amplification & 0.94 & 0.43 & 1.00 & 0.70 & 0.10 & 1.00 & 10 & 293 & 303 \\
Primary & NBN amplification & 0.94 & 0.48 & 1.00 & 0.66 & 0.06 & 1.00 & 8 & 376 & 384 \\
Primary & APC oncogenic mutation & 0.92 & 0.41 & 1.00 & 0.54 & 0.06 & 1.00 & 10 & 374 & 384 \\
Primary & RECQL4 amplification & 0.91 & 0.40 & 0.89 & 0.72 & 0.07 & 1.00 & 9 & 375 & 384 \\
Primary & AGO2 amplification & 0.90 & 0.49 & 0.86 & 0.76 & 0.08 & 1.00 & 7 & 296 & 303 \\
Primary & TMB-H & 0.90 & 0.39 & 1.00 & 0.68 & 0.07 & 1.00 & 9 & 368 & 377 \\
\bottomrule
\end{tabular}

}
\end{subtable}
\end{table}

\begin{table}\ContinuedFloat
\begin{subtable}[h]{\linewidth}
\caption{Biomarkers detected in Renal Cell Carcinoma}\label{table:results-renal-cell-carcinoma}\centering\resizebox{\linewidth}{!}{\begin{tabular}{llrrrrrrrrr}
\toprule
Tumor Site & Biomarker & AUC & AP & Sensitivity & Specificity & PPV & NPV & Positive & Negative & Total \\
\midrule
Primary & TSC2 oncogenic alterations: deletion/mutation & 0.99 & 0.71 & 1.00 & 0.93 & 0.29 & 1.00 & 5 & 168 & 173 \\
Primary & CRLF2 deletion & 0.98 & 0.57 & 1.00 & 0.93 & 0.27 & 1.00 & 4 & 169 & 173 \\
Primary & TSC2 oncogenic mutation & 0.98 & 0.70 & 1.00 & 0.90 & 0.24 & 1.00 & 5 & 168 & 173 \\
Primary & CDKN2A (p14ARF) deletion & 0.96 & 0.55 & 0.83 & 0.92 & 0.28 & 0.99 & 6 & 167 & 173 \\
Primary & CDKN2A (p16INK4a) deletion & 0.96 & 0.53 & 0.83 & 0.92 & 0.26 & 0.99 & 6 & 167 & 173 \\
Primary & CDKN2B oncogenic alterations: deletion/mutation & 0.95 & 0.52 & 0.80 & 0.90 & 0.20 & 0.99 & 5 & 168 & 173 \\
Primary & VHL oncogenic mutation & 0.87 & 0.78 & 0.97 & 0.72 & 0.70 & 0.97 & 69 & 104 & 173 \\
\bottomrule
\end{tabular}

}
\end{subtable}

\begin{subtable}[h]{\linewidth}
\caption{Biomarkers detected in Thyroid Cancer}\label{table:results-thyroid-cancer}\centering\resizebox{\linewidth}{!}{\begin{tabular}{llrrrrrrrrr}
\toprule
Tumor Site & Biomarker & AUC & AP & Sensitivity & Specificity & PPV & NPV & Positive & Negative & Total \\
\midrule
Primary & RB1 oncogenic alterations: deletion/mutation & 0.99 & 0.58 & 1.00 & 0.78 & 0.12 & 1.00 & 2 & 68 & 70 \\
Primary & RB1 oncogenic mutation & 0.97 & 0.45 & 1.00 & 0.93 & 0.29 & 1.00 & 2 & 68 & 70 \\
Primary & MSI-H/dMMR & 0.97 & 0.67 & 1.00 & 0.83 & 0.17 & 1.00 & 2 & 59 & 61 \\
Primary & MSI-H vs MSS & 0.95 & 0.33 & 1.00 & 0.90 & 0.25 & 1.00 & 2 & 59 & 61 \\
Primary & TP53 oncogenic alterations: deletion/mutation & 0.91 & 0.83 & 0.92 & 0.60 & 0.34 & 0.97 & 13 & 57 & 70 \\
Primary & BRAF oncogenic mutation & 0.91 & 0.93 & 0.88 & 0.58 & 0.64 & 0.85 & 32 & 38 & 70 \\
Primary & TMB-H & 0.91 & 0.19 & 1.00 & 0.84 & 0.15 & 1.00 & 2 & 68 & 70 \\
\bottomrule
\end{tabular}

}
\end{subtable}
\end{table}

\begin{table}\ContinuedFloat
\begin{subtable}[h]{\linewidth}
\caption{Biomarkers detected in Soft Tissue Sarcoma}\label{table:results-soft-tissue-sarcoma}\centering\resizebox{\linewidth}{!}{\begin{tabular}{llrrrrrrrrr}
\toprule
Tumor Site & Biomarker & AUC & AP & Sensitivity & Specificity & PPV & NPV & Positive & Negative & Total \\
\midrule
Primary & WT1 fusion & 0.99 & 0.83 & 0.90 & 0.97 & 0.43 & 1.00 & 10 & 346 & 356 \\
Primary & SDHC amplification & 0.94 & 0.29 & 1.00 & 0.47 & 0.05 & 1.00 & 10 & 346 & 356 \\
Primary & TMB-H & 0.89 & 0.19 & 0.88 & 0.66 & 0.06 & 1.00 & 8 & 338 & 346 \\
Primary & JUN oncogenic alterations: amplification/mutation & 0.89 & 0.39 & 1.00 & 0.58 & 0.06 & 1.00 & 10 & 346 & 356 \\
Primary & RB1 deletion & 0.89 & 0.22 & 0.92 & 0.73 & 0.11 & 1.00 & 13 & 343 & 356 \\
Primary & CYSLTR2 deletion & 0.89 & 0.26 & 0.92 & 0.70 & 0.11 & 1.00 & 13 & 306 & 319 \\
Primary & MDM2 oncogenic alterations: amplification/mutation & 0.88 & 0.76 & 0.85 & 0.70 & 0.44 & 0.94 & 78 & 278 & 356 \\
\bottomrule
\end{tabular}

}
\end{subtable}
\end{table}

\begin{table}
\caption{Performance Metrics for Biomarker Prediction on the TCGA validation cohort. Metrics include Area Under the Curve (AUC), Average Precision (AP), Sensitivity, Specificity, Positive Predictive Value (PPV), and Negative Predictive Value (NPV). Sample counts are provided for each biomarker, indicating the number of positive, negative, and total samples. The performance of biomarkers in 20 TCGA projects that are associated with 12 out of the 15 most common cancer types are summarized in the following sub-tables \ref{table:tcga-results-bladder-cancer} - \ref{table:tcga-results-thyroid-cancer}.} Biomarkers associated with ovarian cancer (TCGA-OV), pancreatic cancer (TCGA-PAAD), and melanoma (TCGA-SKCM, TCGA-UVM) did not pass the criteria for high-performing biomarkers.
\label{table:tcga-validation}
\begin{subtable}[h]{\linewidth}
\caption{Biomarkers detected in Bladder Cancer (TCGA-BLCA)}\label{table:tcga-results-bladder-cancer}
\centering\resizebox{\linewidth}{!}{\begin{tabular}{lrrrrrrrrr}
\toprule
Biomarker & AUC & AP & Sensitivity & Specificity & PPV & NPV & Positive & Negative & Total \\
\midrule
H3C12 amplification & 0.98 & 0.56 & 1.00 & 0.33 & 0.06 & 1.00 & 3 & 76 & 79 \\
H3C2 amplification & 0.98 & 0.56 & 1.00 & 0.21 & 0.05 & 1.00 & 3 & 76 & 79 \\
H3C10 amplification & 0.97 & 0.67 & 1.00 & 0.29 & 0.05 & 1.00 & 3 & 76 & 79 \\
H3C7 amplification & 0.96 & 0.61 & 1.00 & 0.26 & 0.05 & 1.00 & 3 & 76 & 79 \\
H3C4 amplification & 0.96 & 0.59 & 1.00 & 0.22 & 0.05 & 1.00 & 3 & 76 & 79 \\
H3C8 amplification & 0.95 & 0.34 & 1.00 & 0.26 & 0.05 & 1.00 & 3 & 76 & 79 \\
MDC1 amplification & 0.94 & 0.32 & 1.00 & 0.72 & 0.12 & 1.00 & 3 & 76 & 79 \\
\bottomrule
\end{tabular}

}
\end{subtable}
\end{table}

\begin{table}\ContinuedFloat
\begin{subtable}[h]{\linewidth}
\caption{Biomarkers detected in Breast Cancer (TCGA-BRCA)}\label{table:tcga-results-breast-cancer}
\centering\resizebox{\linewidth}{!}{\begin{tabular}{lrrrrrrrrr}
\toprule
Biomarker & AUC & AP & Sensitivity & Specificity & PPV & NPV & Positive & Negative & Total \\
\midrule
CDKN1A amplification & 0.99 & 0.59 & 1.00 & 0.29 & 0.03 & 1.00 & 4 & 186 & 190 \\
PIM1 amplification & 0.96 & 0.29 & 1.00 & 0.42 & 0.04 & 1.00 & 5 & 185 & 190 \\
ER & 0.92 & 0.97 & 0.93 & 0.79 & 0.93 & 0.77 & 147 & 47 & 194 \\
FBXW7 oncogenic mutation & 0.90 & 0.13 & 1.00 & 0.20 & 0.03 & 1.00 & 4 & 184 & 188 \\
CDH1 oncogenic alterations: deletion/mutation & 0.90 & 0.51 & 0.85 & 0.86 & 0.42 & 0.98 & 20 & 165 & 185 \\
TP53 oncogenic alterations: deletion/mutation & 0.89 & 0.80 & 0.89 & 0.71 & 0.64 & 0.91 & 70 & 119 & 189 \\
TP53 oncogenic mutation & 0.88 & 0.79 & 0.90 & 0.71 & 0.64 & 0.92 & 70 & 119 & 189 \\
\bottomrule
\end{tabular}

}
\end{subtable}
\end{table}

\begin{table}\ContinuedFloat
\begin{subtable}[h]{\linewidth}
\caption{Biomarkers detected in Colorectal Cancer (TCGA-COAD, TCGA-READ)}\label{table:tcga-results-colorectal-cancer}
\centering\resizebox{\linewidth}{!}{\begin{tabular}{lrrrrrrrrr}
\toprule
Biomarker & AUC & AP & Sensitivity & Specificity & PPV & NPV & Positive & Negative & Total \\
\midrule
BLM oncogenic mutation & 0.99 & 0.58 & 1.00 & 0.42 & 0.04 & 1.00 & 2 & 83 & 85 \\
BLM oncogenic alterations: deletion/mutation & 0.99 & 0.58 & 1.00 & 0.49 & 0.05 & 1.00 & 2 & 81 & 83 \\
ZFHX3 oncogenic mutation & 0.99 & 0.75 & 1.00 & 0.75 & 0.09 & 1.00 & 2 & 80 & 82 \\
KMT2D oncogenic mutation & 0.98 & 0.45 & 1.00 & 0.90 & 0.20 & 1.00 & 2 & 81 & 83 \\
HLA-B deletion/mutation & 0.97 & 0.65 & 1.00 & 0.63 & 0.11 & 1.00 & 4 & 84 & 88 \\
ZFHX3 oncogenic alterations: deletion/mutation & 0.97 & 0.64 & 1.00 & 0.59 & 0.06 & 1.00 & 2 & 80 & 82 \\
MSI-H/dMMR & 0.97 & 0.92 & 0.88 & 0.86 & 0.56 & 0.97 & 16 & 81 & 97 \\
\bottomrule
\end{tabular}

}
\end{subtable}
\end{table}

\begin{table}\ContinuedFloat
\begin{subtable}[h]{\linewidth}
\caption{Biomarkers detected in Endometrial Cancer (TCGA-UCEC, TCGA-UCS)}\label{table:tcga-results-endometrial-cancer}
\centering\resizebox{\linewidth}{!}{\begin{tabular}{lrrrrrrrrr}
\toprule
Biomarker & AUC & AP & Sensitivity & Specificity & PPV & NPV & Positive & Negative & Total \\
\midrule
EZH2 oncogenic mutation & 1.00 & 1.00 & 1.00 & 0.85 & 0.13 & 1.00 & 2 & 84 & 86 \\
PALB2 oncogenic alterations: deletion/mutation & 0.99 & 0.83 & 1.00 & 0.81 & 0.11 & 1.00 & 2 & 84 & 86 \\
EZH2 oncogenic alterations: amplification/mutation & 0.99 & 0.58 & 1.00 & 0.67 & 0.07 & 1.00 & 2 & 81 & 83 \\
KDM5C oncogenic alterations: deletion/mutation & 0.99 & 0.75 & 1.00 & 0.36 & 0.04 & 1.00 & 2 & 77 & 79 \\
ERCC3 oncogenic mutation & 0.98 & 0.50 & 1.00 & 0.60 & 0.06 & 1.00 & 2 & 86 & 88 \\
RET oncogenic mutation & 0.98 & 0.50 & 1.00 & 0.76 & 0.09 & 1.00 & 2 & 82 & 84 \\
EZH2 oncogenic alterations: deletion/mutation & 0.98 & 0.70 & 1.00 & 0.78 & 0.10 & 1.00 & 2 & 81 & 83 \\
\end{tabular}

}
\end{subtable}
\end{table}

\begin{table}\ContinuedFloat
\addtocounter{subtable}{-1}
\begin{subtable}[h]{\linewidth}
\caption{Biomarkers detected in Endometrial Cancer (TCGA-UCEC, TCGA-UCS) (continued)}
\centering\resizebox{\linewidth}{!}{\begin{tabular}{lrrrrrrrrr}
Biomarker & AUC & AP & Sensitivity & Specificity & PPV & NPV & Positive & Negative & Total \\
\midrule
TMB-H & 0.87 & 0.88 & 0.79 & 0.84 & 0.80 & 0.83 & 42 & 51 & 93 \\
TRIP13 amplification & 0.87 & 0.11 & 1.00 & 0.41 & 0.04 & 1.00 & 2 & 90 & 92 \\
ERCC4 oncogenic alterations: deletion/mutation & 0.87 & 0.13 & 1.00 & 0.69 & 0.08 & 1.00 & 2 & 78 & 80 \\
NSD1 oncogenic mutation & 0.87 & 0.35 & 1.00 & 0.62 & 0.26 & 1.00 & 10 & 74 & 84 \\
SRC amplification & 0.87 & 0.38 & 0.75 & 0.80 & 0.14 & 0.99 & 4 & 88 & 92 \\
POLE oncogenic mutation & 0.87 & 0.42 & 1.00 & 0.55 & 0.24 & 1.00 & 11 & 78 & 89 \\
TERT amplification & 0.87 & 0.13 & 1.00 & 0.73 & 0.08 & 1.00 & 2 & 90 & 92 \\
\end{tabular}

}
\end{subtable}
\end{table}

\begin{table}\ContinuedFloat
\addtocounter{subtable}{-1}
\begin{subtable}[h]{\linewidth}
\caption{Biomarkers detected in Endometrial Cancer (TCGA-UCEC, TCGA-UCS) (continued)}
\centering\resizebox{\linewidth}{!}{\begin{tabular}{lrrrrrrrrr}
Biomarker & AUC & AP & Sensitivity & Specificity & PPV & NPV & Positive & Negative & Total \\
\midrule
TP63 amplification & 0.80 & 0.15 & 1.00 & 0.66 & 0.15 & 1.00 & 5 & 86 & 91 \\
PRDM14 amplification & 0.80 & 0.10 & 1.00 & 0.51 & 0.06 & 1.00 & 3 & 89 & 92 \\
IL7R amplification & 0.80 & 0.13 & 1.00 & 0.62 & 0.08 & 1.00 & 3 & 89 & 92 \\
PTCH1 oncogenic mutation & 0.80 & 0.33 & 0.82 & 0.68 & 0.26 & 0.96 & 11 & 77 & 88 \\
PPM1D amplification & 0.80 & 0.07 & 1.00 & 0.41 & 0.04 & 1.00 & 2 & 90 & 92 \\
EIF4A2 amplification & 0.80 & 0.14 & 1.00 & 0.71 & 0.17 & 1.00 & 5 & 86 & 91 \\
MAP3K13 amplification & 0.80 & 0.20 & 0.75 & 0.78 & 0.25 & 0.97 & 8 & 83 & 91 \\
\bottomrule
\end{tabular}

}
\end{subtable}
\end{table}

\begin{table}\ContinuedFloat
\begin{subtable}[h]{\linewidth}
\caption{Biomarkers detected in Esophagogastric Cancer (TCGA-ESCA, TCGA-STAD)}\label{table:tcga-results-esophagogastric-cancer}
\centering\resizebox{\linewidth}{!}{\begin{tabular}{lrrrrrrrrr}
\toprule
Biomarker & AUC & AP & Sensitivity & Specificity & PPV & NPV & Positive & Negative & Total \\
\midrule
BMPR1A oncogenic mutation & 0.99 & 0.58 & 1.00 & 0.38 & 0.03 & 1.00 & 2 & 91 & 93 \\
B2M oncogenic mutation & 0.99 & 0.59 & 1.00 & 0.94 & 0.38 & 1.00 & 3 & 90 & 93 \\
MEN1 oncogenic alterations: deletion/mutation & 0.98 & 0.48 & 1.00 & 0.22 & 0.04 & 1.00 & 3 & 88 & 91 \\
B2M oncogenic alterations: deletion/mutation & 0.97 & 0.44 & 1.00 & 0.83 & 0.17 & 1.00 & 3 & 89 & 92 \\
PIK3C2G amplification & 0.97 & 0.42 & 1.00 & 0.29 & 0.03 & 1.00 & 2 & 91 & 93 \\
ARID1B oncogenic mutation & 0.97 & 0.66 & 1.00 & 0.47 & 0.06 & 1.00 & 3 & 89 & 92 \\
CDH1 oncogenic mutation & 0.96 & 0.56 & 1.00 & 0.84 & 0.22 & 1.00 & 4 & 86 & 90 \\
\bottomrule
\end{tabular}

}
\end{subtable}
\end{table}

\begin{table}\ContinuedFloat 
\begin{subtable}[h]{\linewidth}
\caption{Biomarkers detected in Glioma (TCGA-GBM, TCGA-LGG)}\label{table:tcga-results-glioma}
\centering\resizebox{\linewidth}{!}{\begin{tabular}{lrrrrrrrrr}
\toprule
Biomarker & AUC & AP & Sensitivity & Specificity & PPV & NPV & Positive & Negative & Total \\
\midrule
GRIN2A oncogenic alterations: deletion/mutation & 0.97 & 0.56 & 1.00 & 0.68 & 0.07 & 1.00 & 3 & 136 & 139 \\
IDH1 oncogenic mutation & 0.94 & 0.92 & 0.91 & 0.81 & 0.84 & 0.89 & 76 & 68 & 144 \\
IDH1 oncogenic alterations: amplification/mutation & 0.93 & 0.93 & 0.92 & 0.76 & 0.81 & 0.89 & 76 & 67 & 143 \\
TP53 oncogenic alterations: deletion/mutation & 0.93 & 0.93 & 0.95 & 0.64 & 0.66 & 0.95 & 60 & 83 & 143 \\
TP53 oncogenic mutation & 0.93 & 0.92 & 0.95 & 0.64 & 0.64 & 0.95 & 58 & 86 & 144 \\
XRCC2 amplification & 0.93 & 0.29 & 1.00 & 0.49 & 0.07 & 1.00 & 6 & 166 & 172 \\
FUBP1 oncogenic alterations: deletion/mutation & 0.93 & 0.48 & 1.00 & 0.54 & 0.10 & 1.00 & 7 & 134 & 141 \\
\bottomrule
\end{tabular}

}
\end{subtable}
\end{table}

\begin{table}\ContinuedFloat
\begin{subtable}[h]{\linewidth}
\caption{Biomarkers detected in Hepatobiliary Cancer (TCGA-CHOL, TCGA-LIHC)}\label{table:tcga-results-hepatobiliary-cancer}
\centering\resizebox{\linewidth}{!}{\begin{tabular}{lrrrrrrrrr}
\toprule
Biomarker & AUC & AP & Sensitivity & Specificity & PPV & NPV & Positive & Negative & Total \\
\midrule
DNAJB1 fusion & 1.00 & 1.00 & 1.00 & 0.99 & 0.67 & 1.00 & 2 & 88 & 90 \\
FGFR2 oncogenic alterations: amplification/mutation & 0.99 & 0.58 & 1.00 & 0.87 & 0.15 & 1.00 & 2 & 85 & 87 \\
FGFR oncogenic mutation & 0.95 & 0.36 & 1.00 & 0.89 & 0.17 & 1.00 & 2 & 88 & 90 \\
ALOX12B deletion & 0.93 & 0.57 & 1.00 & 0.29 & 0.03 & 1.00 & 2 & 86 & 88 \\
FLCN deletion & 0.93 & 0.19 & 1.00 & 0.52 & 0.05 & 1.00 & 2 & 85 & 87 \\
IDH1 oncogenic alterations: amplification/mutation & 0.92 & 0.58 & 0.80 & 0.95 & 0.50 & 0.99 & 5 & 82 & 87 \\
GPS2 deletion & 0.88 & 0.17 & 1.00 & 0.36 & 0.04 & 1.00 & 2 & 86 & 88 \\
\bottomrule
\end{tabular}

}
\end{subtable}
\end{table}

\begin{table}\ContinuedFloat
\begin{subtable}[h]{\linewidth}
\caption{Biomarkers detected in Non-Small Cell Lung Cancer (TCGA-LUAD, TCGA-LUSC)}\label{table:tcga-results-non-small-cell-lung-cancer}
\centering\resizebox{\linewidth}{!}{\begin{tabular}{lrrrrrrrrr}
\toprule
Biomarker & AUC & AP & Sensitivity & Specificity & PPV & NPV & Positive & Negative & Total \\
\midrule
STK11 oncogenic alterations: deletion/mutation & 0.94 & 0.51 & 1.00 & 0.64 & 0.16 & 1.00 & 12 & 169 & 181 \\
GNAS amplification & 0.94 & 0.19 & 1.00 & 0.28 & 0.03 & 1.00 & 4 & 179 & 183 \\
STK11 oncogenic mutation & 0.94 & 0.46 & 1.00 & 0.75 & 0.22 & 1.00 & 12 & 169 & 181 \\
CDK4 oncogenic alterations: amplification/mutation & 0.87 & 0.19 & 1.00 & 0.52 & 0.06 & 1.00 & 5 & 176 & 181 \\
CDK4 amplification & 0.87 & 0.28 & 1.00 & 0.53 & 0.06 & 1.00 & 5 & 178 & 183 \\
RTEL1 amplification & 0.87 & 0.29 & 0.83 & 0.43 & 0.05 & 0.99 & 6 & 177 & 183 \\
KRAS oncogenic mutation & 0.86 & 0.44 & 0.96 & 0.67 & 0.29 & 0.99 & 23 & 159 & 182 \\
\bottomrule
\end{tabular}

}
\end{subtable}
\end{table}

\begin{table}\ContinuedFloat
\begin{subtable}[h]{\linewidth}
\caption{Biomarkers detected in Prostate Cancer (TCGA-PRAD)}\label{table:tcga-results-prostate-cancer}
\centering\resizebox{\linewidth}{!}{\begin{tabular}{lrrrrrrrrr}
\toprule
Biomarker & AUC & AP & Sensitivity & Specificity & PPV & NPV & Positive & Negative & Total \\
\midrule
FANCC amplification & 0.92 & 0.20 & 1.00 & 0.25 & 0.03 & 1.00 & 2 & 79 & 81 \\
BMPR1A oncogenic alterations: deletion/mutation & 0.85 & 0.13 & 1.00 & 0.30 & 0.04 & 1.00 & 2 & 79 & 81 \\
ERCC3 deletion & 0.85 & 0.17 & 1.00 & 0.40 & 0.08 & 1.00 & 4 & 77 & 81 \\
ZRSR2 amplification & 0.84 & 0.16 & 1.00 & 0.69 & 0.11 & 1.00 & 3 & 78 & 81 \\
PTCH1 amplification & 0.84 & 0.11 & 1.00 & 0.30 & 0.04 & 1.00 & 2 & 79 & 81 \\
MED12 oncogenic alterations: amplification/mutation & 0.81 & 0.09 & 1.00 & 0.60 & 0.06 & 1.00 & 2 & 78 & 80 \\
TBX3 amplification & 0.81 & 0.53 & 1.00 & 0.23 & 0.03 & 1.00 & 2 & 79 & 81 \\
SPOP oncogenic alterations: deletion/mutation & 0.78 & 0.22 & 1.00 & 0.21 & 0.12 & 1.00 & 8 & 73 & 81 \\
KMT2C fusion & 0.76 & 0.11 & 1.00 & 0.38 & 0.04 & 1.00 & 2 & 79 & 81 \\
SPOP oncogenic mutation & 0.76 & 0.20 & 1.00 & 0.33 & 0.14 & 1.00 & 8 & 73 & 81 \\
CDK12 oncogenic alterations: deletion/mutation & 0.75 & 0.22 & 1.00 & 0.43 & 0.07 & 1.00 & 3 & 76 & 79 \\
\bottomrule
\end{tabular}

}
\end{subtable}
\end{table}

\begin{table}\ContinuedFloat
\begin{subtable}[h]{\linewidth}
\caption{Biomarkers detected in Renal Cell Carcinoma (TCGA-KICH, TCGA-KIRC, TCGA-KIRP)}\label{table:tcga-results-renal-cell-carcinoma}
\centering\resizebox{\linewidth}{!}{\begin{tabular}{lrrrrrrrrr}
\toprule
Biomarker & AUC & AP & Sensitivity & Specificity & PPV & NPV & Positive & Negative & Total \\
\midrule
MET oncogenic mutation & 0.97 & 0.53 & 1.00 & 0.69 & 0.07 & 1.00 & 3 & 138 & 141 \\
VHL oncogenic mutation & 0.86 & 0.36 & 0.82 & 0.80 & 0.29 & 0.98 & 11 & 109 & 120 \\
MET oncogenic alterations: amplification/mutation & 0.85 & 0.34 & 0.80 & 0.76 & 0.11 & 0.99 & 5 & 136 & 141 \\
PBRM1 oncogenic alterations: deletion/mutation & 0.85 & 0.51 & 0.90 & 0.71 & 0.37 & 0.98 & 21 & 115 & 136 \\
PBRM1 oncogenic mutation & 0.84 & 0.52 & 0.84 & 0.71 & 0.32 & 0.96 & 19 & 116 & 135 \\
TP53 oncogenic alterations: deletion/mutation & 0.83 & 0.23 & 0.80 & 0.78 & 0.12 & 0.99 & 5 & 137 & 142 \\
BRCA1/2 germline/somatic mutation/deletion or fusion & 0.82 & 0.11 & 0.75 & 0.79 & 0.08 & 0.99 & 4 & 163 & 167 \\
TP53 oncogenic mutation & 0.82 & 0.17 & 0.80 & 0.80 & 0.12 & 0.99 & 5 & 137 & 142 \\
DNMT3A oncogenic alterations: deletion/mutation & 0.79 & 0.22 & 0.75 & 0.60 & 0.05 & 0.99 & 4 & 138 & 142 \\
KRAS oncogenic mutation & 0.78 & 0.37 & 1.00 & 0.42 & 0.04 & 1.00 & 3 & 139 & 142 \\
FGFR4 amplification & 0.78 & 0.18 & 1.00 & 0.35 & 0.09 & 1.00 & 10 & 157 & 167 \\
FGFR4 oncogenic alterations: amplification/mutation & 0.78 & 0.17 & 1.00 & 0.40 & 0.11 & 1.00 & 10 & 131 & 141 \\
\bottomrule
\end{tabular}

}
\end{subtable}
\end{table}

\begin{table}\ContinuedFloat
\begin{subtable}[h]{\linewidth}
\caption{Biomarkers detected in Soft Tissue Sarcoma (TCGA-SARC)}\label{table:tcga-results-soft-tissue-sarcoma}
\centering\resizebox{\linewidth}{!}{\begin{tabular}{lrrrrrrrrr}
\toprule
Biomarker & AUC & AP & Sensitivity & Specificity & PPV & NPV & Positive & Negative & Total \\
\midrule
KLF4 amplification & 1.00 & 1.00 & 1.00 & 0.53 & 0.08 & 1.00 & 2 & 49 & 51 \\
GNB1 amplification & 0.96 & 0.45 & 1.00 & 0.37 & 0.06 & 1.00 & 2 & 49 & 51 \\
TNFRSF14 amplification & 0.95 & 0.37 & 1.00 & 0.29 & 0.05 & 1.00 & 2 & 49 & 51 \\
ROS1 oncogenic alterations: amplification/mutation & 0.95 & 0.64 & 1.00 & 0.88 & 0.25 & 1.00 & 2 & 48 & 50 \\
RAC1 oncogenic alterations: amplification/mutation & 0.92 & 0.34 & 1.00 & 0.48 & 0.11 & 1.00 & 3 & 48 & 51 \\
IRS1 deletion & 0.92 & 0.27 & 1.00 & 0.71 & 0.12 & 1.00 & 2 & 48 & 50 \\
PIK3CA oncogenic mutation & 0.91 & 0.35 & 1.00 & 0.65 & 0.10 & 1.00 & 2 & 51 & 53 \\
\bottomrule
\end{tabular}

}
\end{subtable}
\end{table}

\begin{table}\ContinuedFloat
\begin{subtable}[h]{\linewidth}
\caption{Biomarkers detected in Thyroid Cancer (TCGA-THCA)}\label{table:tcga-results-thyroid-cancer}
\centering\resizebox{\linewidth}{!}{\begin{tabular}{lrrrrrrrrr}
\toprule
Biomarker & AUC & AP & Sensitivity & Specificity & PPV & NPV & Positive & Negative & Total \\
\midrule
NRAS oncogenic alterations: amplification/mutation & 0.94 & 0.43 & 1.00 & 0.75 & 0.28 & 1.00 & 8 & 83 & 91 \\
BRAF oncogenic mutation & 0.93 & 0.94 & 0.96 & 0.40 & 0.65 & 0.89 & 49 & 42 & 91 \\
HRAS oncogenic mutation & 0.92 & 0.44 & 0.75 & 0.87 & 0.21 & 0.99 & 4 & 87 & 91 \\
BRAF oncogenic alterations: amplification/mutation & 0.92 & 0.94 & 0.96 & 0.46 & 0.69 & 0.90 & 50 & 41 & 91 \\
NRAS oncogenic mutation & 0.91 & 0.36 & 1.00 & 0.78 & 0.31 & 1.00 & 8 & 83 & 91 \\
HRAS oncogenic alterations: amplification/mutation & 0.88 & 0.27 & 0.75 & 0.86 & 0.20 & 0.99 & 4 & 87 & 91 \\
\bottomrule
\end{tabular}

}
\end{subtable}
\end{table}

\input{tables/msk-counts}  

\clearpage
\begin{landscape}
\begingroup
\renewcommand{\arraystretch}{1.05}
\begin{table}[!ht]
\caption{TCGA Development cohort counts}
\label{table:tcga-counts}

\centering\resizebox{\linewidth}{!}{
\begin{tabular}{llrrrrrr}
\toprule
 &  & \multicolumn{3}{r}{train} & \multicolumn{3}{r}{test} \\
 &  & patients & samples & slides & patients & samples & slides \\
Cancer Type & TCGA Project &  &  &  &  &  &  \\
\midrule
Breast Cancer & TCGA-BRCA & 835 & 835 & 894 & 205 & 205 & 214 \\
Non-Small Cell Lung Cancer & TCGA-LUAD, TCGA-LUSC & 741 & 741 & 822 & 204 & 204 & 219 \\
Glioma & TCGA-GBM, TCGA-LGG & 687 & 687 & 1299 & 187 & 187 & 389 \\
Renal Cell Carcinoma & TCGA-KICH, TCGA-KIRC, TCGA-KIRP & 654 & 654 & 680 & 180 & 180 & 182 \\
Endometrial Cancer & TCGA-UCEC, TCGA-UCS & 433 & 433 & 506 & 95 & 95 & 108 \\
Colorectal Cancer & TCGA-COAD, TCGA-READ & 423 & 423 & 424 & 99 & 99 & 100 \\
Esophagogastric Cancer & TCGA-ESCA, TCGA-STAD & 422 & 422 & 445 & 95 & 95 & 97 \\
Thyroid Cancer & TCGA-THCA & 407 & 407 & 418 & 91 & 91 & 92 \\
Head and Neck Cancer & TCGA-HNSC & 366 & 366 & 381 & 80 & 80 & 87 \\
Prostate Cancer & TCGA-PRAD & 317 & 317 & 356 & 83 & 83 & 90 \\
Hepatobiliary Cancer & TCGA-CHOL, TCGA-LIHC & 307 & 307 & 314 & 94 & 94 & 95 \\
Bladder Cancer & TCGA-BLCA & 306 & 306 & 374 & 79 & 79 & 82 \\
Cervical Cancer & TCGA-CESC & 201 & 201 & 211 & 61 & 61 & 61 \\
Soft Tissue Sarcoma & TCGA-SARC & 195 & 195 & 470 & 53 & 53 & 119 \\
Miscellaneous Neuroepithelial Tumor & TCGA-PCPG & 145 & 145 & 163 & 31 & 31 & 31 \\
Pancreatic Cancer & TCGA-PAAD & 142 & 142 & 156 & 40 & 40 & 46 \\
Melanoma & TCGA-SKCM, TCGA-UVM & 122 & 122 & 123 & 57 & 57 & 69 \\
Germ Cell Tumor & TCGA-TGCT & 104 & 104 & 183 & 45 & 45 & 62 \\
Thymic Epithelial Tumor & TCGA-THYM & 97 & 97 & 144 & 24 & 24 & 36 \\
Ovarian Cancer & TCGA-OV & 79 & 79 & 80 & 27 & 27 & 27 \\
Mesothelioma & TCGA-MESO & 59 & 59 & 70 & 16 & 16 & 17 \\
Adrenocortical Carcinoma & TCGA-ACC & 44 & 44 & 179 & 11 & 11 & 47 \\
Mature B-Cell Neoplasms & TCGA-DLBC & 33 & 33 & 33 & 11 & 11 & 11 \\
\bottomrule
\end{tabular}
}
\end{table}
\endgroup
\end{landscape}

\clearpage
\subsection{Supplementary Figures}
\setcounter{figure}{0}
\renewcommand{\thefigure}{S\arabic{figure}}

\begin{figure}[!ht]
    \centering
    \includegraphics[width=\textwidth]{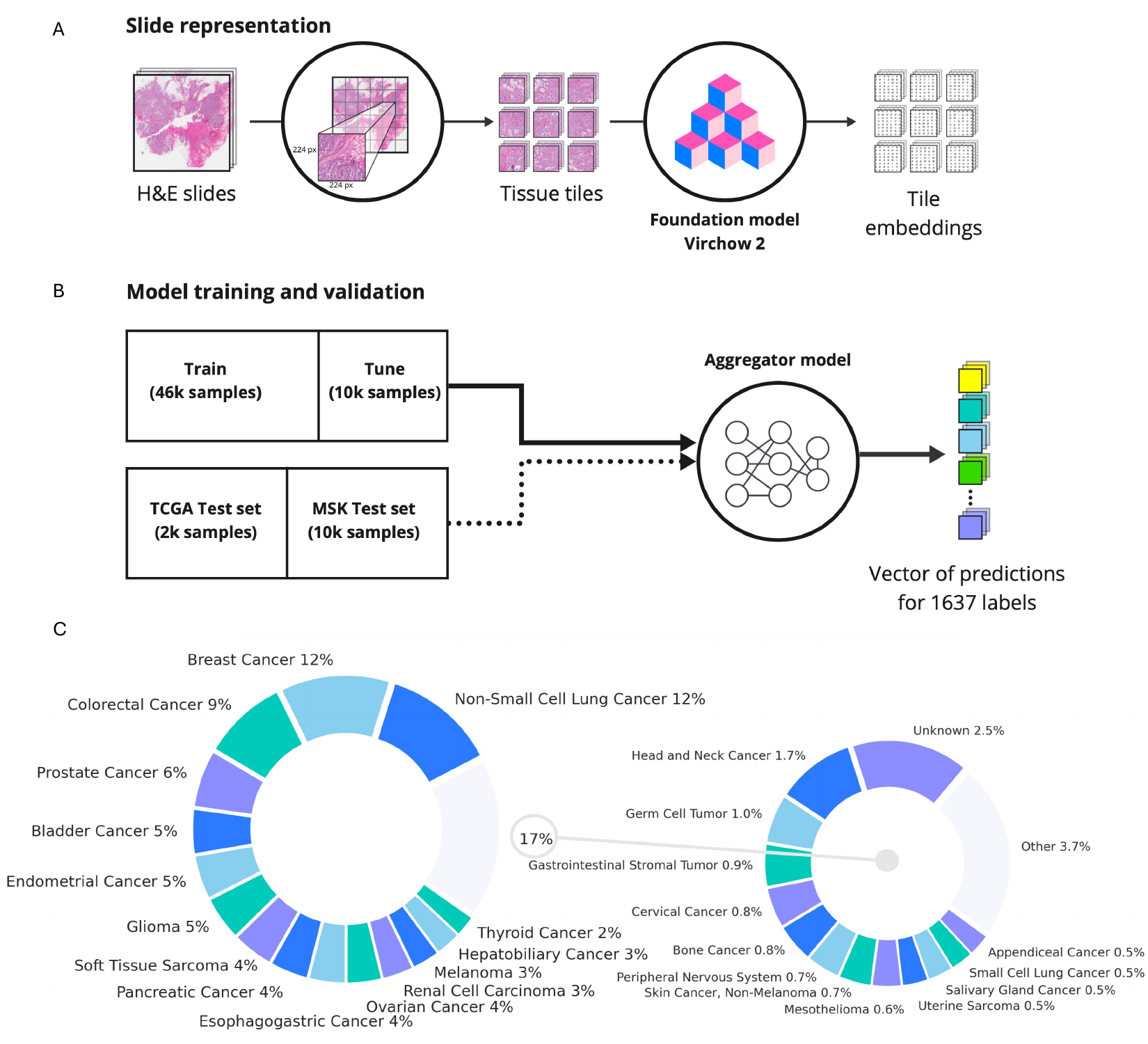}
    \caption{
    \textbf{A}: Schematic diagram of slide representation generation. The image is split into tiles of size $224\times224$ pixels. The tiles are filtered such that only those representing tissue are kept. Next, the tissue tiles are fed to Virchow2, which generates an embedding of length 1,280 for each tile. Each slide is thus represented as a collection of its tissue tile embeddings.
    \textbf{B}: Schematic flow chart of model training and validation. The aggregator model receives the tile embeddings of the WSIs in the train set (processed as described in \textbf{A}) and runs validation during training on the tune set. Once trained, the model is evaluated on two unseen test sets: the TCGA and MSK cohorts.
    \textbf{C}: Donut charts showing the distribution of histologies in the MSK dataset. The left chart displays the 15 most common cancer types, representing 83\% of all samples. The right chart shows the composition of cancer types in the remaining 17\% of samples, including those in the `Unknown' category and rare cancer samples in the `Other' category. 
    }
    \label{fig:S1}
\end{figure}

\begin{figure}[ht]
    \centering
    \includegraphics[width=\textwidth]{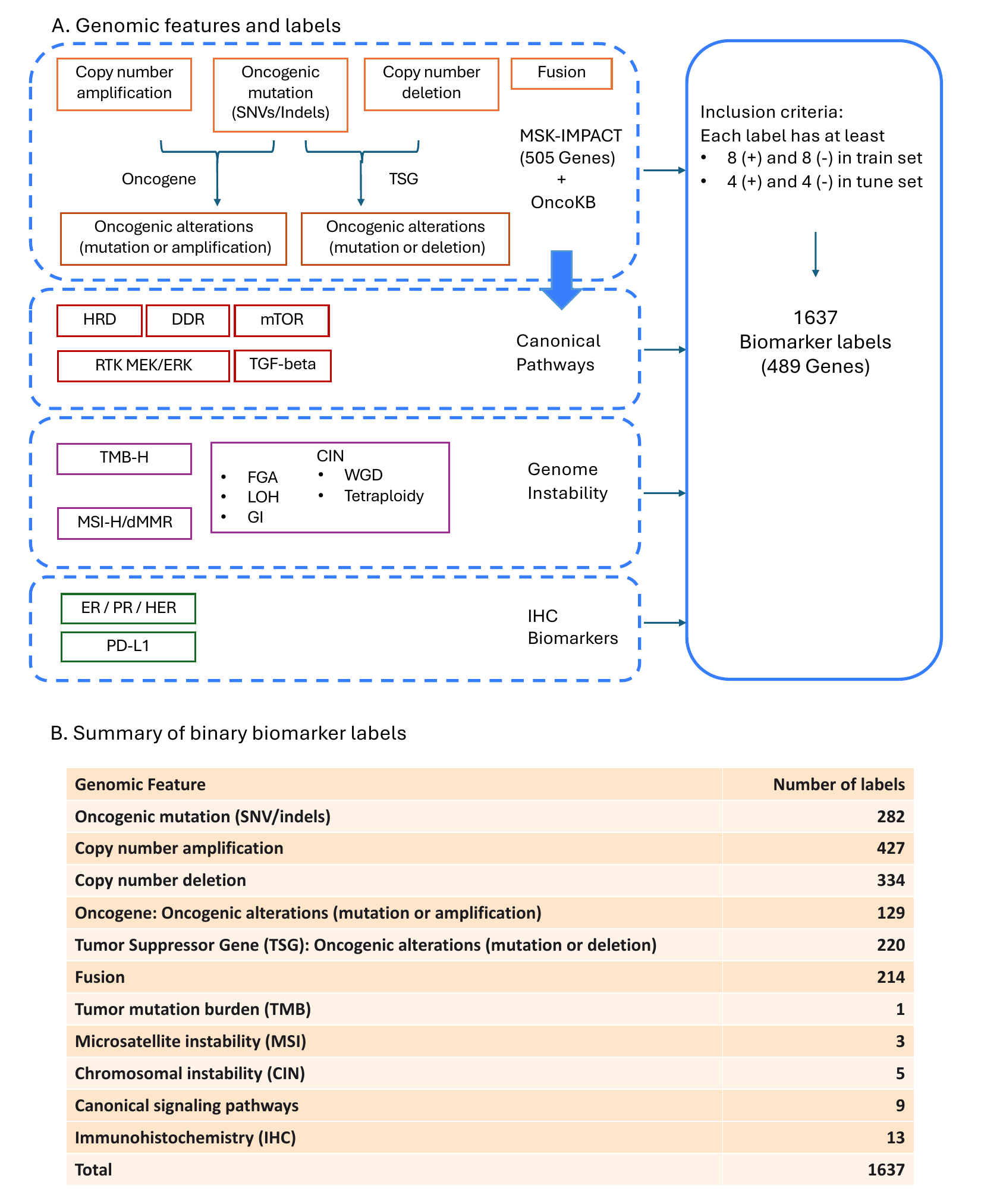}
    \caption{
    Genomic feature and binary biomarker label derivation. 
    \textbf{A}: Overview of genomic features, including gene level alteration labels of \nimpactgenes{} genes from the MSK-IMPACT panel, alterations in a group of genes participating in 5 canonical signaling pathways, genome instability: tumor mutation burden (TMB), microsatellite instability high (MSI-H) or defects in mismatch repair genes (dMMR), and \acf{CIN} measured by fraction of genome altered (FGA), loss-of-heterozygosity (LOH), genome instability index (GI), whole genome doubling (WGD), and tetraploidy. The alterations include mutations (SNVs/indels), copy number aberrations (amplification and deletion), and fusion events. The oncogenic status was determined based on OncoKB annotation. 
    \textbf{B}: Summary table showing the number of labels in different label categories. 
    }
    \label{fig:S2}
\end{figure}

\begin{figure}
    \centering
    \includegraphics[width=\textwidth]{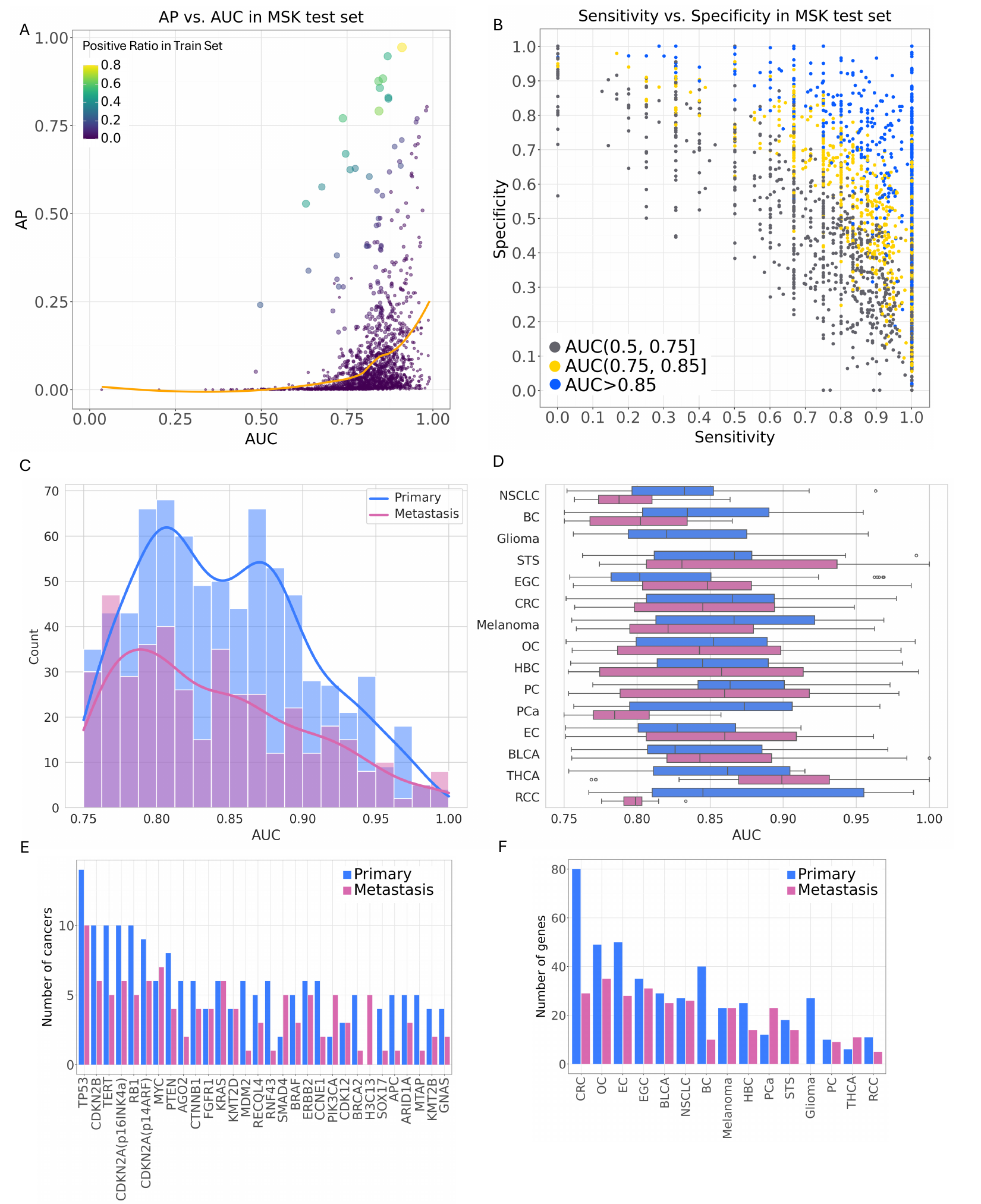}
    \caption{
    Gene biomarkers identified across MSK common cancers. 
    \textbf{A}: Scatter plot showing the average precision (AP) vs. AUC of biomarkers obtained in MSK test set. Color and size indicating the proportion of positive samples in the train set. 
    \textbf{B}: Scatter plot showing sensitivity vs. specificity of biomarkers obtained in MSK test set. Color coded by the ranges of AUC:~$(0.5,0.75]$, $(0.75, 0.85]$, and $(0.85,1.0]$.
    \textbf{C}: Histogram showing the distribution of the AUC score across all cancer types, comparing primary and metastatic samples. \textbf{D}:~Boxplots showing the distribution of the AUC score in each of the 15 common cancer types, comparing primary and metastatic samples.
    \textbf{E}:~Top 30 gene biomarkers identified across 15 most common cancers, with AUC\,$>0.75$. 
    \textbf{F}:~Distribution of number of genes biomarkers identified across 15 most common cancers.}
    \label{fig:S3}
\end{figure}

\begin{figure}
    \centering
    \includegraphics[width=\textwidth]{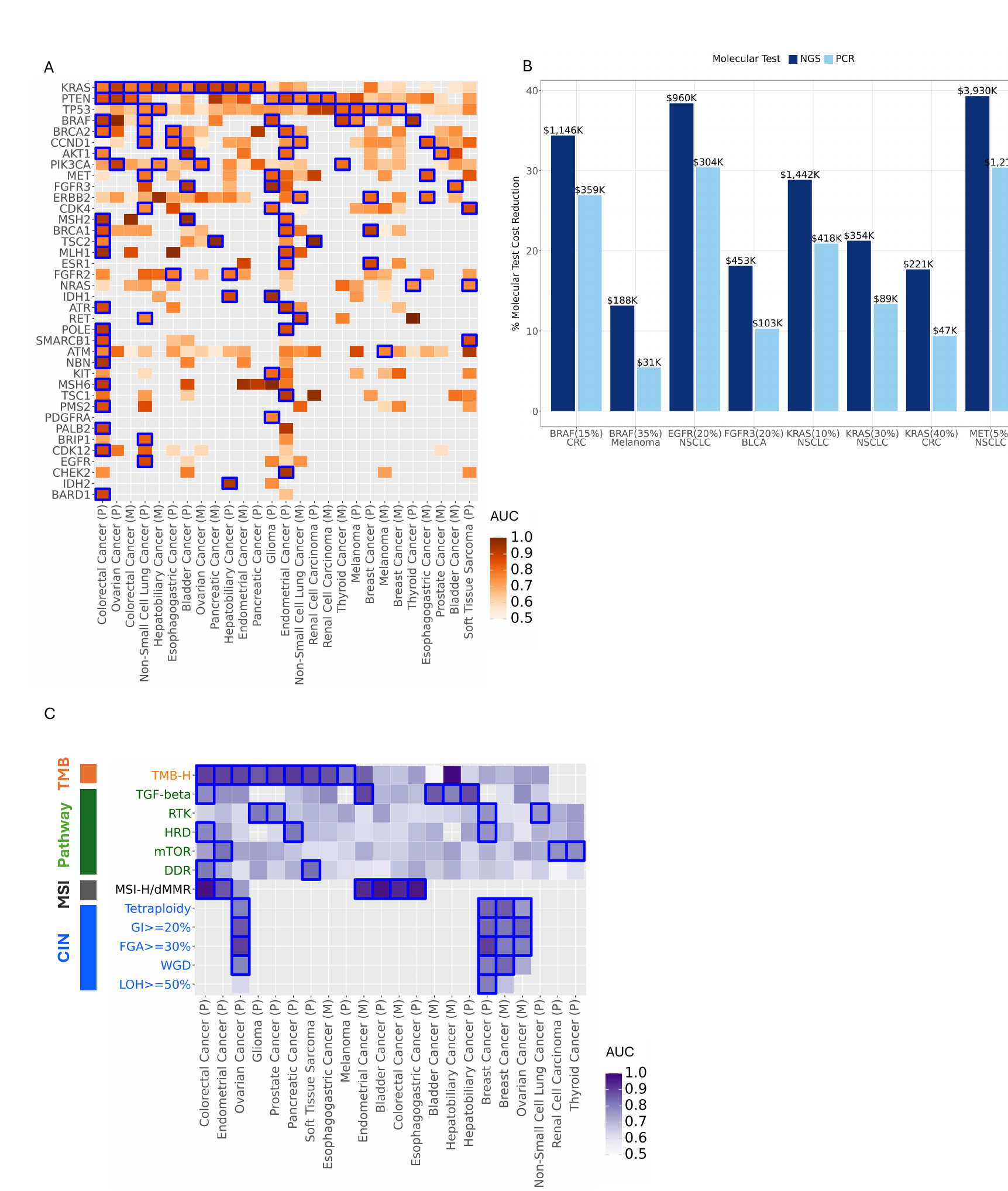}
    \caption{
    Biomarker predictions in treatment targets and pathways. 
    \textbf{A}:~Heatmap showing the prediction of specific therapeutic hotspots in target genes (Y-axis) detected by OmniScreen across common cancer types at primary (P) and metastatic (M) lesions, with AUC~$>0.5$. 
    Blue boxes indicate high-performing therapeutic targets associated biomarkers (Table~\ref{table:filter-criteria}).
    \textbf{B}: Estimated cost saving for PCR or NGS definitive screening test, using OmniScreen pre-screening, to enroll 500 patients with a mutation in a targeted gene and histology. Y-axis showed the estimated cost reduction (\%) for molecular testing with OmniScreen pre-screening. The estimated amount~(\$) of cost reduction is annotated on top of each bar. X-axis indicated targeted genes that are commonly assessed by NGS or PCR clinically, and their published prevalence of gene alterations (\%) in each cancer: BRAF in colorectal cancer (CRC) and melanoma, EGFR in non-small cell lung cancer (NSCLC), FGFR3 in bladder cancer, 
    KRAS (estimated prevalence: 10\% in non-smokers, 30\% in smokers for NSCLC; 40\% in CRC), 
    and MET in NSCLC.
    \textbf{C}: Prediction of 5 canonical signaling pathways (TGF-$\beta$, RTK, HRD, mTOR, DDR), \acf{TMB}, \acf{MSI}, and chromosomal instability measures: genome instability index (GI), fraction of genome altered (FGA), tetraploidy, whole genome doubling (WGD) and loss-of-heterozygosity (LOH). Blue boxes indicate high-performing biomarkers (Table~\ref{table:filter-criteria}).
    }
    \label{fig:S4}
\end{figure}

\begin{figure}
    \centering
    \includegraphics[width=\textwidth]{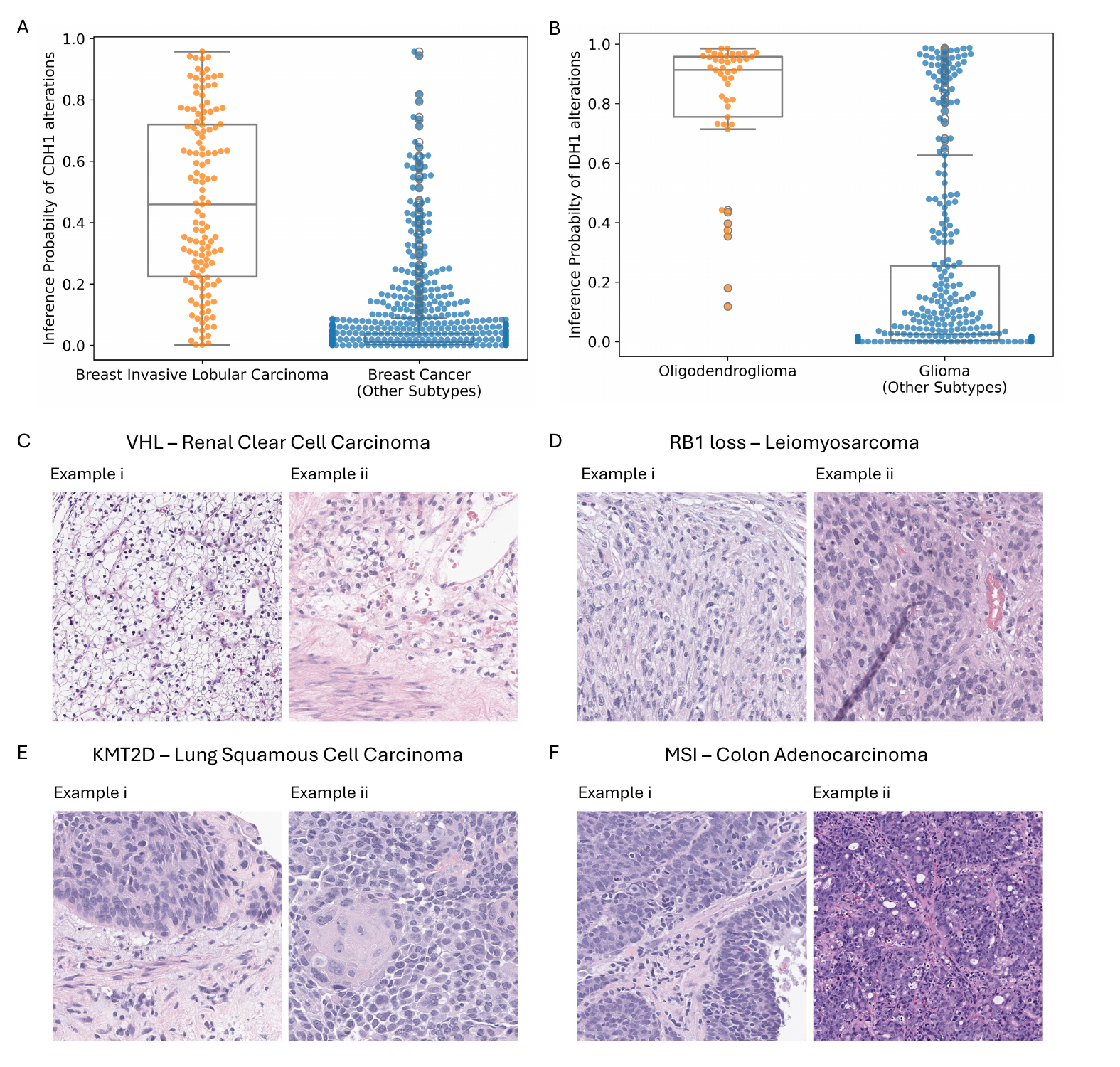}
    \caption{
    Genotype-Phenotype associations.
    \textbf{A}:~Comparison between the distribution of CDH1 prediction scores in breast lobular invasive carcinoma and in the other subtypes of breast cancers.
    \textbf{B}:~Comparison between the distribution of IDH1 prediction scores in oligodendroglioma and in the other subtypes of glioma.
    Examples of $1024\times1024$ regions highly attended by OmniScreen during the prediction of genomic alterations, highlighting specific histologic phenotypic features. These include 
    \textbf{C}:~
    cases with VHL mutations confirmed as clear cell renal cell carcinoma, displaying the characteristic clear cell histology of renal cell carcinoma ((i) and (ii)); 
    \textbf{D}:~cases with RB1 loss identified as leiomyosarcomas, showing the characteristic spindle cells with eosinophilic cytoplasm in storiform or fasicular growth patterns characteristic of leiomyosarcomas (i), with perhaps higher concentration of nuclear atypia (ii). 
    \textbf{E}:~Cases with KMT2D mutations detected as lung squamous cell carcinomas, exhibiting moderate to high nuclear pleomorphism and atypia (i) along with features characteristic of squamous cell differentiation, such as intercellular bridges and focal keratinization (ii);  
    \textbf{F}:~colon adenocarcinomas with MSI-H/dMMR, characterized by poorly differentiated, medullary-like patterns of tumor growth with ``pushing'' borders, tumor infiltrating lymphocytes (TILs) ((i) and (ii)), and mucin production. 
    }
    \label{fig:S5}
\end{figure}

\clearpage
\bibliographystyle{unsrt}
\bibliography{references}

\begin{acronym}[MSK-IMPACT]
    \acro{ADH}{atypical ductal hyperplasia}
    \acro{ADT}{androgen deprivation therapy}
    \acro{AI}{artificial intelligence}
    \acro{ALH}{atypical lobular hyperplasia}
    \acro{AP}{average precision}
    \acro{AR}{androgen receptor}
    \acro{AUROC}[AUC]{area under the receiver operating characteristic curve}
    \acro{BLN}{Breast Lymph Node}
    \acro{BRAF}{B-Raf proto-oncogene}
    \acro{ccRCC}{clear cell renal cell carcinoma}
    \acro{CDH1}{cadherin 1}
    \acro{consep}[CoNSeP]{colorectal nuclear segmentation and phenotypes}
    \acro{CIN}{chromosomal instability}
    \acro{CRC}{colorectal cancer}
    \acro{CRPC}{castration resistant prostate cancer}
    \acro{DCIS}{ductal carcinoma in situ}
    \acro{DDLPS}{dedifferentiated liposarcomas}
    \acro{DLBCL}{diffused large B-cell lymphoma}
    \acro{dMMR}{deficient mismatch repair}
    \acro{DDR}{DNA damage response}
    \acro{EGFR}{epidermal growth factor receptor}
    \acro{ER}{estrogen receptor}
    \acro{EMA}{exponential moving average}
    \acro{FCN}{Fully Convolutional Network}
    \acro{FDA}{Food and Drug Administration}
    \acro{FGA}{fraction of genome altered}
    \acro{FGFR}{fibroblast growth factor receptor}
    \acro{FISH}{Fluorescence In Situ Hybridization}
    \acro{FL}{follicular lymphoma}
    \acro{FN}{false negative}
    \acro{FPR}{false positive rate}
    \acro{FP}{false positive}
    \acro{GELU}{Gaussian Error Linear Unit}
    \acro{GI}{gastrointestinal}
    \acro{GI index}{genome instability index}
    \acro{HER2}{human epidermal growth factor receptor 2}
    \acro{HE}[H\&E]{hematoxylin and eosin}
    \acro{HGSOC}{high-grade serous ovarian cancer}
    \acro{HIPT}{hierarchical image pyramid transformer}
    \acro{HN}[H\&N]{head and neck}
    \acro{HRD}{homologous recombination deﬁciency}
    \acro{iBOT}{image BERT pre-training with online tokenizer}
    \acro{ICI}{immune checkpoint inhibitor}
    \acro{ID}{in-distribution}
    \acro{IDC}{invasive ductal carcinoma}
    \acro{IHC}{immunohistochemistry}
    \acro{IMPACT}{Integrated Mutation Profiling of Actionable Cancer Targets}
    \acro{ILC}{invasive lobular carcinoma}
    \acro{indels}{insertions/deletions}
    \acro{ITC}{infiltrating tumor cell}
    \acro{KS}{Kolmogorov–Smirnov}
    \acro{LCIS}{lobular carcinoma in situ}
    \acro{LUSC}{lung squamous cell carcinoma}
    \acro{LOH}{loss-of-heterozygosity}
    \acro{LS}{Lynch Syndrome}
    \acro{MAE}{masked autoencoder}
    \acro{MCG}{My Cancer Genome}
    \acro{MIL}{multiple instance learning}
    \acro{MMR}{mismatch repair}
    \acro{mpp}{microns-per-pixel}
    \acro{MSI-H}{microsatellite instability high}
    \acro{MSI}{microsatellite instability}
    \acro{MSK}{Memorial Sloan Kettering}
    \acro{MSK-IMPACT}{Memorial Sloan Kettering-Integrated Mutation Profiling of Actionable Targets}
    \acro{MSKCC}{Memorial Sloan Kettering Cancer Center}
    \acro{MTC}{medullary thyroid cancer}
    \acro{MSS}{microsatellite stable}
    \acro{MZL}{marginal zone lymphoma}
    \acro{NCCN}{National Comprehensive Cancer Network}
    \acro{NCI}{National Cancer Institute}
    \acro{NGS}{next-generation sequencing}
    \acro{NPV}{negative predictive value}
    \acro{NSCLC}{non-small cell lung cancer}
    \acro{OOD}{out-of-distribution}
    \acro{OCCC}{ovarian clear cell carcinoma}
    \acro{PAIP}{Pathology AI Platform}
    \acro{PARP}{poly(ADP-ribose) polymerase}
    \acro{PCam}{PatchCamelyon}
    \acro{PCA}{principal component analysis}
    \acro{PCR}{polymerase chain reaction}
    \acro{PD-L1}{programmed death-ligand 1}
    \acro{PR}{progesterone receptor}
    \acro{PTEN}{phosphatase and tensin homolog}
    \acro{RCT}{randomised controlled trial}
    \acro{RET}{ret proto-oncogene}
    \acro{ROC}{receiver operating characteristic}
    \acro{RCC}{renal cell carcinoma}
    \acro{RTK}{tyrosine kinase receptor}
    \acro{SGD}{stochastic gradient descent}
    \acro{SNVs} {single-nucleotide variants}
    \acro{TCGA}{The Cancer Genome Atlas}
    \acro{TIL}{tumor infiltrating lymphocyte}
    \acro{TKI}{tyrosine kinase inhibitor}
    \acro{TMB}{tumor mutation burden}
    \acro{TMB-H}{tumor mutation burden-high}
    \acro{TNR}{true negative rate}
    \acro{TN}{true negative}
    \acro{TPR}{true positive rate}
    \acro{TP}{true positive}
    \acro{TP53}{tumor protein p53}
    \acro{TSG}{tumor suppressor gene}
    \acro{ViT}{vision transformer}
    \acro{WDLPS}{well-differentiated liposarcomas}
    \acro{WGD}{whole genome doubling}
    \acro{WSI}{whole slide image}
\end{acronym}

\end{document}